\documentclass[sigconf]{acmart}

\usepackage[utf8]{inputenc}
\usepackage{listings} %
\usepackage[linesnumbered,ruled,vlined]{algorithm2e}
\usepackage{forest}
\forestset{
    default preamble={
        for tree = {
            font=\footnotesize    %
        }
    }
}

\usepackage{fancybox}

\usepackage{booktabs}
\usepackage{soul}
\usepackage{tablefootnote}
\usepackage{graphicx} %
\graphicspath{{images/}} %

\usepackage{amsthm}

\usepackage{balance}

\usepackage{blindtext}
\usepackage{tabularx}

\usepackage{pgfplots}
\pgfplotsset{width=7cm,compat=1.18}
\usepgfplotslibrary{groupplots}
\usetikzlibrary{patterns}
\usepackage{pgfplotstable}

\newcommand\examplesCombined{\ensuremath{S_1}}
\newcommand\examplesRecurse{\ensuremath{S_1}}   %

\newcommand\grammar{\ensuremath{\mathcal{G}}}

\newcommand{\term}[1]{\textcolor{teal}{\texttt{#1}}}

\newcommand{\boxalign}[2][0.9\columnwidth]{
  \par\noindent\tikzstyle{mybox} = [draw=black,inner sep=6pt]
  \begin{center}\begin{tikzpicture}
   \node [mybox] (box){%
    \begin{minipage}{#1}{#2}\end{minipage}
   };
  \end{tikzpicture}\end{center}
}

\newcommand{\glade}{\textsc{Glade}}
\newcommand{\arvada}{\textsc{Arvada}}
\newcommand{\toolName}{\textsc{TreeVada}}

\def\BibTeX{{\rm B\kern-.05em{\sc i\kern-.025em b}\kern-.08em
    T\kern-.1667em\lower.7ex\hbox{E}\kern-.125emX}}

\copyrightyear{2024}
\acmYear{2024}
\setcopyright{rightsretained}
\acmConference[ICSE '24]{2024 IEEE/ACM 46th International Conference on
Software Engineering}{April 14--20, 2024}{Lisbon, Portugal}
\acmBooktitle{2024 IEEE/ACM 46th International Conference on Software
Engineering (ICSE '24), April 14--20, 2024, Lisbon,
Portugal}\acmDOI{10.1145/3597503.3639214}
\acmISBN{979-8-4007-0217-4/24/04}

\begin{document}

\author{Mohammad Rifat Arefin} %
\affiliation{%
  \department{Computer Science and Engineering Department}
  \institution{University of Texas at Arlington}
  \streetaddress{500 UTA Blvd}
  \city{Arlington}
  \state{Texas}
  \postcode{76019}
  \country{USA}
}

\author{Suraj Shetiya} 
\affiliation{%
  \department{Computer Science and Engineering Department}
  \institution{University of Texas at Arlington}
  \streetaddress{500 UTA Blvd}
  \city{Arlington}
  \state{Texas}
  \postcode{76019}
  \country{USA}
}

\author{Zili Wang}
\affiliation{%
  \department{Department of Computer Science}
  \institution{Iowa State University}
  \streetaddress{2433 Union Dr.}
  \city{Ames}
  \state{Iowa}
  \postcode{50011}
  \country{USA}
}

\author{Christoph Csallner}
\affiliation{%
  \department{Computer Science and Engineering Department}
  \institution{University of Texas at Arlington}
  \streetaddress{500 UTA Blvd}
  \city{Arlington}
  \state{Texas}
  \postcode{76019}
  \country{USA}
}

\title{Fast Deterministic Black-box Context-free Grammar Inference}

\begin{abstract}
Black-box context-free grammar inference is a hard problem as in many practical settings it only has access to a limited number of example programs. The state-of-the-art approach \arvada{} heuristically generalizes grammar rules starting from flat parse trees and is non-deterministic to explore different generalization sequences. We observe that many of \arvada{}'s generalization steps violate common language concept nesting rules. We thus propose to pre-structure input programs along these nesting rules, apply learnt rules recursively, and make black-box context-free grammar inference deterministic. The resulting \toolName{} yielded faster runtime and higher-quality grammars in an empirical comparison. The \toolName{} source code, scripts, evaluation parameters, and training data are open-source and publicly available (\url{https://doi.org/10.6084/m9.figshare.23907738}). 
\end{abstract}



\maketitle

\section{Introduction}

Learning a context-free grammar from sample programs with just the help of a black-box parser currently does not scale well to realistic settings. Existing approaches need a combination of a large number of sample programs (deep learning), the ability to manipulate a grey-box or white-box parser, or are non-deterministic. The most closely related approach, the recent \arvada{} work~\cite{arvada21ASE}, is non-deterministic and thus ran \arvada{} 10~times for each input to explore different sequences of grammar inference steps.

Black-box context-free grammar inference is crucially important when a language only has a black-box parser that cannot be instrumented. On the other hand, program samples are often available (e.g., as open-source code or as example programs from the language vendor). Examples of such languages typically only have closed-source parsers that are often only available remotely (or cannot be instrumented for legal reasons). This unfortunately rules out using white-box or grey-box parser instrumentation~\cite{Lin2010TSE,Autogram2016ASE,REINAM2019FSE,gopinath20mining}. 

At the same time, grammar inference is important for many software engineering tasks, including code comprehension~\cite{oda2015learning}, reverse engineering~\cite{moonen2001generating}, detecting and refactoring code smells~\cite{kim2013specification, nagy2017static}, transforming source code~\cite{adhikari2021simulink} for optimization or bug fixing, and generating test inputs~\cite{srivastava2021gramatron,nguyen2020mofuzz,alsaeed2023finding, godefroid2008grammar}.

The task of black-box inference of a context-free grammar is fundamentally hard. First, the given input programs likely do not cover all aspects of their language's ``golden grammar''. Second, it is often very hard to generalize from a few programs exhibiting a few combinations of language features to a grammar describing the language features with the correct nesting rules. Finally, not being able to inspect or instrument the language's parser makes black-box inference significantly harder than grey-box or white-box inference, as a black-box approach has a much narrower access to the parser's encoding of the language's golden grammar.

While there has been a lot of interest in applying deep learning techniques to learning grammars from program samples~\cite{LearnFuzz2017ASE,DeepSmith2018ISSTA}, a principal limitation of deep-learning approaches is that (a)~they need a very large amount of training samples (which may not be available) and (b)~they do not take advantage of black-box parsers that are typically available even for closed-source languages. Indeed, the \arvada{} paper reports on a comparison with state-of-the-art deep learning approaches, in which deep-learning tools did not match the precision of either \glade{}~\cite{Glade17PLDI} or \arvada{}~\cite{arvada21ASE}.

While \arvada{}~\cite{arvada21ASE} has made significant improvements over the pioneering \glade{}~\cite{Glade17PLDI,Glade22Replication} work, it still has several limitations. For example, \arvada{} has $O(n^4)$ runtime in its $n$ input tokens and requires its ``seed'' input programs to be very short. For example, on average \arvada{} mostly produced~\cite{arvada21ASE} grammars within 5~minutes with over 80\%~F1~scores when running on a few dozen hand-selected minimal sample programs that on average consist of just 1.7 to 12.5~characters. However when running on 25~randomly generated nodejs programs with an average length of 50~characters, \arvada{} yielded on average a 29\%~F1~score---after a 12~hour runtime.

\toolName{} combines several new techniques. First, to guide its grammar generalization steps to avoid breaking common nesting rules, \toolName{} first pre-structures its input programs according to nesting rules induced by balanced brackets that are common in many languages~\cite{van2019lightweight}. Depending on the language grammar's nesting structure this step reduces \toolName{}'s runtime from \arvada{}'s $O(n^4)$ downward up to $O(n^2)$. Second, once \toolName{} accepts a grammar generalization step, \toolName{} applies this generalization rule recursively. Finally, building on these techniques \toolName{} carefully omits non-determinism and thus yields a reproducible grammar in a single run. In an empirical comparison the resulting \toolName{} implementation achieved both faster runtime and better grammar quality than the most-closely-related \arvada{} tool.
To summarize, the paper makes the following major contributions.
\begin{itemize}
    \item \toolName{} is the first fast deterministic black-box approach for context-free grammar inference that produces high-quality grammars.
    \item The paper compares \toolName{} empirically with its closest competitor (\arvada{}) using \arvada{}'s setup and achieves faster runtime and better grammar quality.
    \item The \toolName{} source code, scripts, evaluation parameters, and training data are open-source and publicly available (\textbf{\url{https://github.com/rifatarefin/treevada}}) 
    and archived (\url{https://doi.org/10.6084/m9.figshare.23907738}).
\end{itemize}

\begin{figure}
        \centering
    \boxalign{{ 
        \begin{math}
        \textit{start}  \to~ \textit{stmt}\\
        \textit{stmt}  \to~ \textit{stmt}~\term{\textvisiblespace{};\textvisiblespace{}}~\textit{stmt}~ 
        |~\term{L\textvisiblespace{}=\textvisiblespace{}}~\textit{numexpr}~
        |~\term{skip}\\        |~\term{while\textvisiblespace}~\textit{boolexpr}~\term{\textvisiblespace{}do\textvisiblespace{}}~ \textit{stmt}\\
        |~\term{if\textvisiblespace{}}~\textit{boolexpr}~\term{\textvisiblespace{}then\textvisiblespace{}}~ \textit{stmt} ~\term{\textvisiblespace{}else\textvisiblespace{}}~ \textit{stmt}\\
        \textit{boolexpr}  \to~  \term{$\sim$}\textit{boolexpr}~|~\term{true}~|~\term{false}\\
        |~\textit{boolexpr}~\term{\textvisiblespace{}\&\textvisiblespace{}}~\textit{boolexpr}~      
        |~\textit{numexpr}~\term{\textvisiblespace{}==\textvisiblespace{}}~\textit{numexpr}\\
        \textit{numexpr}  \to~ \term{(}\,\textit{numexpr}~\term{+}~\textit{numexpr}\,\term{)}~|~\term{L}~|~\term{n}
        \end{math}
    }
    }
    \caption{\textit{while}'s golden grammar $\grammar_w$
    (Arvada's motivating example~\cite[Figure~1]{arvada21ASE}, reformatted, plus missing skip rule).}
    \label{fig:while}
\end{figure}

\section{Background}

While there are trade-offs and special cases, ``context-free'' 
remains an important abstraction level for programming language definition, both for human-level programming language understanding and for automated language processing tools. For example, the latest versions of the official language specifications of complex mainstream languages such as Java~\cite{gosling2022java}, JavaScript~\cite{ECMAscript2022}, and C++~\cite{cplusplus} include in their language syntax descriptions context-free grammars. Similarly, many sample grammars of the widely-used ANTLR4 line of parser generators~\cite{parr2014adaptive} are context-free.

Figure~\ref{fig:while} is a small example context-free grammar. As usual, a rule with alternatives ($l \to r_1 \dots r_m$~|~$r_n \dots r_z$) is just shorthand for having both a first ($l \to r_1 \dots r_m$) and second ($l \to r_n \dots r_z$) rule. Each rule is thus essentially of the same form ($l \to r_1 \dots r_m$) with a single non-terminal ($l$) on the left and a sequence ($r_1,\dots,r_m$) of terminals, non-terminals, or both on the right. This, of course, allows \emph{recursive rules}
(e.g.: $boolexpr \to$ \term{$\sim$} $boolexpr$) and \emph{balanced nesting} structures (e.g.: $numexpr \to \term{(}\,numexpr\,\term{+}\,numexpr\,\term{)}$).

Many programming languages allow balanced nesting of language concepts, where a concept has a dedicated start terminal and a dedicated end terminal
and a concept can contain other concepts~\cite{van2019lightweight}. For these start and end terminals many languages use matching round, square, and curly brackets: \term{(} \term{)} \term{[} \term{]} \term{\{} \term{\}}. Common thereby nested concepts include
class and function definitions,
parameter lists,
code blocks,
array creation and access,
and various other expressions---for example, a code block containing other code blocks that in turn contain arithmetic expressions.

\subsection{Black-box Grammar Inference}

The long-term goal of this line of work is to reverse engineer the (unknown) grammar (or specification) of a programming language from only two things, (1)~a few valid sample programs and (2)~a black-box parser. Having such a specification would support various software engineering tasks, including 
code comprehension~\cite{oda2015learning},
reverse engineering~\cite{moonen2001generating},
smell detection and refactoring~\cite{kim2013specification,nagy2017static},
test input generation~\cite{hodovan2018grammarinator}, and code transformation~\cite{adhikari2021simulink}.

For example, some popular commercial languages (e.g., MATLAB/Simulink)  neither have a formal specification nor parsers that can be analyzed.
Specifically, the language's tools are closed-source and cannot be instrumented for legal or technical reasons (e.g., they are only available as a remote service). But valid sample programs are often widely available---via GitHub or the vendor's website (to document language features and encourage language adoption).

We follow \arvada{}'s definition~\cite{arvada21ASE} of grammar quality. So an inferred grammar $\grammar_i$ is better if it has a higher F1 score, i.e., if the set of input programs $\grammar_i$ accepts is closer to the set of input programs accepted by the golden grammar.

\subsection{State-of-the-art Inference: \arvada{} in $O(n^4)$}

Black-box inference of context-free grammars was pioneered by \glade{}~\cite{Glade17PLDI,Glade22Replication} and recently advanced by \arvada{}~\cite{arvada21ASE}. \arvada{}'s evaluation showed \arvada{}'s average run provided on average some 5$\times$ improvement in recall and 3$\times$ improvement in F1~score over \glade{} (while being just 30\% slower).

\arvada{} initially treats each input program as a flat parse tree, i.e., a single rule that can only reproduce the given input program. \arvada{} then iteratively generalizes grammar rules. It groups a few (leaf and/or internal) parse-tree sibling nodes 
$t_s,\dots,t_u$ (aka a ``bubble'') under a new bubble parent node $b$, reflecting a candidate grammar rule 
($b\to~t_s \dots t_u)$. 
\arvada{} then picks an existing interior tree node 
$a$ and its child nodes $t_c,\dots,t_e$, which together implicitly define a grammar rule
($a\to~t_c \dots t_e)$.
It then checks if it can rename the new bubble parent node to (and thereby merge it with) the existing interior node, yielding the generalized rule 
($a~\to~t_c \dots t_e$~$|$~$t_s \dots t_u$).

\arvada{} heuristically accepts such a rule merge when a black-box parser accepts up to 100~freshly generated sample programs that exercise the newly-merged rule. Since each such merge check (and especially failing it) is expensive, \arvada{} orders its potential bubbles via heuristics. The key heuristic is to compare the $k$ siblings immediately before (``left $k$-context'') and after (``right $k$-context'') a candidate bubble. \arvada{} thus ranks a bubble higher if the bubble's contexts are more similar to the contexts of existing interior tree nodes. The secondary bubble-ranking metric is each bubble's occurrence count in the input programs' parse trees (a higher bubble occurrence frequency yields a higher rank). To increase the chance of a merge, \arvada{} also tries to merge two bubbles directly with each other (``2-bubble''), for which it ranks all bubble pairs.

The evaluation of \arvada{} (and \glade{})~\cite{arvada21ASE} points to two scalability issues. 
(1)~First, \glade{}'s and \arvada{}'s training sets only contain very small input programs, i.e., the largest input programs range from just 5~(arith language) to 245~characters (tinyc).
(2)~Second, relatively more complex languages (tinyc and especially nodejs) have the relatively larger golden grammars and input programs. Besides the higher runtime, here \glade{} and \arvada{} also yield lower F1~scores.
Following are the key technical challenges of \arvada{}.

\subsubsection{Arvada Run = 10 Non-deterministic $O(n^4)$ Runs}

\begin{figure}[h!t]
\small
{
	\centering    
	\examplesCombined{} = \{ \term{while\textvisiblespace{}n\textvisiblespace{}==\textvisiblespace{}(n+n)\textvisiblespace{}do\textvisiblespace{}L\textvisiblespace{}=\textvisiblespace{}n} , \\
    \term{L\textvisiblespace{}=\textvisiblespace{}((n+n)+n)\textvisiblespace{};\textvisiblespace{}skip} \}
    
}
\resizebox{\columnwidth}{!}{
	\begin{forest} [$t_0$, l sep={1cm}, for tree= {s sep={0.0mm}} 
	[\texttt{while}] 
	[\textvisiblespace{}]
	[\texttt{n}]
	[\textvisiblespace{}]
    [\texttt{=}][\texttt{=}]
    [\textvisiblespace{}]
    [\texttt{(}][\texttt{n}][\texttt{+}][\texttt{n}][\texttt{)}]
    [\textvisiblespace{}]
	[\texttt{do}]
	[\textvisiblespace{}]
	[\texttt{L}]
	[\textvisiblespace{}]
	[\texttt{=}]
	[\textvisiblespace{}]
	[\texttt{n}]
	]	
	\end{forest} 
 \begin{forest}
	[$t_0$, l sep={1cm}, for tree= {s sep={0.0mm}} 
	    [\texttt{L}] 
	[\textvisiblespace{}]
	[\texttt{=}]
	[\textvisiblespace{}]
    [\texttt{(}][\texttt{(}][\texttt{n}][\texttt{+}][\texttt{n}][\texttt{)}][\texttt{+}][\texttt{n}][\texttt{)}]
	[\textvisiblespace{}]
	[\texttt{;}]
	[\textvisiblespace{}]
	[\texttt{skip}]
	  ]
\end{forest}
	}
 
\vspace{1 em}
\small \textsc{MergeAllValid:} merge \ovalbox{\term{skip}} with \ovalbox{$t_0$} into \ovalbox{$t_0$}
\vspace{-.3em}
$$\downarrow$$
\resizebox{0.95\columnwidth}{!}{
	\begin{forest} [$t_0$, fill = green, l sep={1cm}, for tree= {s sep={0.0mm}} 
	[\texttt{while}] 
	[\textvisiblespace{}]
	[\texttt{n}]
	[\textvisiblespace{}]
    [\texttt{=}][\texttt{=}]
    [\textvisiblespace{}]
    [\texttt{(}][\texttt{n}][\texttt{+}][\texttt{n}][\texttt{)}]
    [\textvisiblespace{}]
	[\texttt{do}]
	[\textvisiblespace{}]
	[\texttt{L}]
	[\textvisiblespace{}]
	[\texttt{=}]
	[\textvisiblespace{}]
	[\texttt{n}]
	]	
	\end{forest} 
 \begin{forest}
	[$t_0$, fill = green, l sep={0.5cm}, for tree= {s sep={0.0mm}} 
	    [\texttt{L}] 
	[\textvisiblespace{}]
	[\texttt{=}]
	[\textvisiblespace{}]
    [\texttt{(}][\texttt{(}][\texttt{n}][\texttt{+}][\texttt{n}][\texttt{)}][\texttt{+}][\texttt{n}][\texttt{)}]
	[\textvisiblespace{}]
	[\texttt{;}]
	[\textvisiblespace{}]
	[$t_0$, fill = green[\texttt{skip}]]
	  ]
\end{forest}
	}
 
\vspace{1em}

{ \small  \ovalbox{$t_1$} $\to$  \ovalbox{\texttt{\term{L}\,\term{\textvisiblespace{}}\,\term{=}\,\term{\textvisiblespace{}}\,\term{n}}}, 
 merge \ovalbox{$t_1$} with \ovalbox{$t_0$} into \ovalbox{$t_0$}
 }
\vspace{-.3em}
$$\downarrow$$
\resizebox{0.95\columnwidth}{!}{
	\begin{forest} [$t_0$, fill = lime, l sep={0.5cm}, for tree= {s sep={0.0mm}} 
	[\texttt{while}] 
	[\textvisiblespace{}]
	[\texttt{n}]
	[\textvisiblespace{}]
    [\texttt{=}][\texttt{=}]
    [\textvisiblespace{}]
    [\texttt{(}][\texttt{n}][\texttt{+}][\texttt{n}][\texttt{)}]
    [\textvisiblespace{}]
	[\texttt{do}]
	[\textvisiblespace{}]
    [$t_0$, fill = lime
	[\texttt{L}]
	[\textvisiblespace{}]
	[\texttt{=}]
	[\textvisiblespace{}]
	[\texttt{n}]]
	]	
	\end{forest} 

 \begin{forest}
	[$t_0$, fill = lime, l sep={0.5cm}, for tree= {s sep={0.0mm}} 
	    [\texttt{L}] 
	[\textvisiblespace{}]
	[\texttt{=}]
	[\textvisiblespace{}]
    [\texttt{(}][\texttt{(}][\texttt{n}][\texttt{+}][\texttt{n}][\texttt{)}][\texttt{+}][\texttt{n}][\texttt{)}]
	[\textvisiblespace{}]
	[\texttt{;}]
	[\textvisiblespace{}]
	[$t_0$, fill = lime[\texttt{skip}]]
	  ]
\end{forest}
	}

\vspace{1em}

{ \small \ovalbox{$t_2$} $\to$  \ovalbox{\texttt{\term{(}\,\term{n}\,\term{+}\,\term{n}\,\term{)}}},
 merge \ovalbox{$t_2$} with \ovalbox{\term{\texttt{n}}}:
 \ovalbox{$t_2$} $\to$ \ovalbox{\texttt{\term{(}$t_2$\term{+}$t_2$\term{)} | \term{n}}}
 }
 
\vspace{-1.5em}
$$\downarrow$$
\resizebox{.95\columnwidth}{!}{
	\begin{forest} [$t_0$, l sep={0.5cm}, for tree= {s sep={0.0mm}} 
	[\texttt{while}] 
	[\textvisiblespace{}]
	[$t_2$, fill = yellow[\texttt{n}]]
	[\textvisiblespace{}]
    [\texttt{=}][\texttt{=}]
    [\textvisiblespace{}]
    [$t_2$, fill = yellow[\texttt{(}][$t_2$, fill=yellow[\texttt{n}]][\texttt{+}][$t_2$, fill=yellow[\texttt{n}]][\texttt{)}]]
    [\textvisiblespace{}]
	[\texttt{do}]
	[\textvisiblespace{}]
    [$t_0$
	[\texttt{L}]
	[\textvisiblespace{}]
	[\texttt{=}]
	[\textvisiblespace{}]
	[$t_2$, fill = yellow[\texttt{n}]]]
	]	
	\end{forest} 

 \begin{forest}
	[$t_0$, l sep={0.5cm}, for tree= {s sep={0.0mm}} 
	    [\texttt{L}] 
	[\textvisiblespace{}]
	[\texttt{=}]
	[\textvisiblespace{}]
    [\texttt{(}]
    [$t_2$, fill = yellow[\texttt{(}][$t_2$, fill = yellow[\texttt{n}]][\texttt{+}][$t_2$, fill = yellow[\texttt{n}]][\texttt{)}]]
    [\texttt{+}][$t_2$, fill = yellow[\texttt{n}]][\texttt{)}]
	[\textvisiblespace{}]
	[\texttt{;}]
	[\textvisiblespace{}]
	[$t_0$[\texttt{skip}]]
	  ]
\end{forest}

	}
 \vspace{1em}
 
\resizebox{0.95\columnwidth}{!}{
{ \small  \ovalbox{$t_3$} $\to$  \ovalbox{\texttt{\term{)}\,\term{\textvisiblespace{}}\,\term{;}\,\term{\textvisiblespace{}}\,$t_0$}},
 partially merge with \ovalbox{\term{)}}: \ovalbox{$t_3$} $\to$ \ovalbox{\texttt{$t_3$\,\term{\textvisiblespace{}}\,\term{;}\,\term{\textvisiblespace{}}\,$t_0$ | \term{)}}
 }}}

 \vspace{-1.5em}
 $$\downarrow$$
\resizebox{.95\columnwidth}{!}{
	\begin{forest} [$t_0$, l sep={0.5cm}, for tree= {s sep={0.0mm}} 
	[\texttt{while}] 
	[\textvisiblespace{}]
	[$t_2$[\texttt{n}]]
	[\textvisiblespace{}]
    [\texttt{=}][\texttt{=}]
    [\textvisiblespace{}]
    [$t_2$[\texttt{(}][$t_2$[\texttt{n}]][\texttt{+}][$t_2$[\texttt{n}]][\texttt{)}]]
    [\textvisiblespace{}]
	[\texttt{do}]
	[\textvisiblespace{}]
    [$t_0$
	[\texttt{L}]
	[\textvisiblespace{}]
	[\texttt{=}]
	[\textvisiblespace{}]
	[$t_2$[\texttt{n}]]]
	]	
	\end{forest} 

 \begin{forest}
	[$t_0$, l sep={0.5cm}, for tree= {s sep={0.0mm}} 
	    [\texttt{L}] 
	[\textvisiblespace{}]
	[\texttt{=}]
	[\textvisiblespace{}]
    [\texttt{(}]
    [$t_2$[\texttt{(}][$t_2$[\texttt{n}]][\texttt{+}][$t_2$[\texttt{n}]][\texttt{)}]]
    [\texttt{+}][$t_2$[\texttt{n}]]
    [$t_3$, fill = orange[$t_3$, fill = orange[\texttt{)}]]
	[\textvisiblespace{}]
	[\texttt{;}]
	[\textvisiblespace{}]
	[$t_0$[\texttt{skip}]]]
	  ]
\end{forest}

	}
 
\small
\begin{align*}
    t_0 \to~& \term{while}\, \term{\textvisiblespace{}}\, t_2\, \term{\textvisiblespace{}}\, \term{=}\, \term{=}\, \term{\textvisiblespace{}}\, t_2\, \term{\textvisiblespace{}}\, \term{do}\, \term{\textvisiblespace{}}\, t_0\, \, |\, \, \term{L}\, \term{\textvisiblespace{}}\, \term{=}\, \term{\textvisiblespace{}}\, \term{(}\, t_2\, \term{+}\, t_2\, t_3\, \, |\, \, \term{L}\term{\textvisiblespace{}}\, \term{=}\, \term{\textvisiblespace{}}\, t_2 \, \, |\, \, \term{skip}\\
    t_2 \to~& \term{(}\, t_2\, \term{+}\, t_2\, \term{)} \, \, |\, \, \term{n}\\
    t_3 \to~& t_3\, \term{\textvisiblespace{}}\, \term{;}\, \term{\textvisiblespace{}}\, t_0\, \, |\, \, \term{)} 
\end{align*}

\caption{Top to bottom: Input \textit{while} programs \examplesRecurse{} and a resulting \arvada{} run: 
initial (pre-tokenized) flat parse trees, 
initial node-pair merges (green),
1\textsuperscript{st} bubble merge (lime),
2\textsuperscript{nd} bubble merge (yellow) without reapplying rule, and 3\textsuperscript{rd} bubble merge (orange) breaking tree nesting; resulting grammar.}
\label{fig:arvada-weaknesses}
\end{figure}
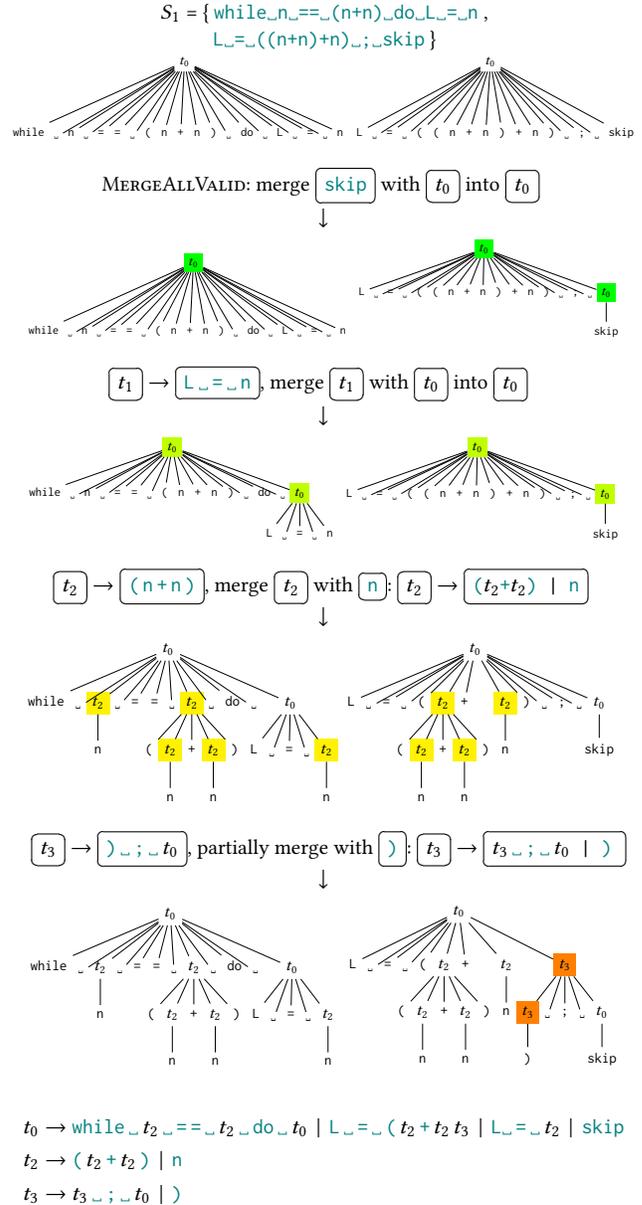

\arvada{}'s first key challenge is its non-determinism, which makes the results hard to reproduce. For example, when we ran it 10~times on the Figure~\ref{fig:arvada-weaknesses} input programs~\examplesRecurse{}, \arvada{} produced two different grammars. Non-determinism also creates a trade-off between using the first run's grammar vs. re-running \arvada{} in the hope of finding a better grammar. On each set of input programs, the \arvada{} work ran \arvada{} 10~times, effectively yielding an order-of-magnitude worse total runtime than the reported average runtime.

\arvada{} uses non-determinism to explore various sequences of grammar generalization steps. Such a generalization sequence can get \arvada{} stuck in the sense of cutting off subsequent generalization options, reducing the inferred grammar's quality. For example, the \arvada{} study reported for several languages a high F1 score variance among its 10~runs. For example, among 10 nodejs runs F1~scores ranged from 0.14 to 0.55. 
Following are \arvada{}'s four main sources of non-determinism.

\textbf{Shuffling Initial Candidate Node-pair Merges:}
\arvada{} first tokenizes each input program along character classes (lower-case, uppercase, digits, whitespace), keeping only 
other ASCII (aka ``punctuation'') and non-ASCII characters as individual tokens. (\arvada{} treats each such resulting token as the only child of a token-specific ``dummy'' parent node connected to the root node---for brevity we omit from figures these dummy nodes.)

\arvada{} then tries an initial attempt (``\textsc{MergeAllValid}'') to generalize grammar rules. Specifically, it creates all pairs of existing non-terminal (mostly dummy) nodes across all parse trees, orders the pairs arbitrarily (by storing them in a non-deterministic data structure), and tries to merge each pair. For example, as Figure~\ref{fig:arvada-weaknesses}'s two initial parse trees contain 12~unique non-terminal node types ($t_0$ and implicit parents of \term{while}, \term{\textvisiblespace{}}, \term{n}, etc.), \arvada{} tries merging 66~node pairs, which yields one successful merge (\term{skip} with $t_0$).

\textbf{Ranking \& Shuffling $O(n^4)$ Candidate Merges:}
After the initialization phase, for each grammar generalization step \arvada{} first (re-) collects and (re-) ranks all possible parse-tree sibling-node sequences (``1-bubbles'') up to a configurable length together with their pairs (``2-bubble''). For $n$ tokens in the initial parse trees there are essentially $O(n^2)$ 1-bubbles, which makes the ranking overall $O(n^4)$. \arvada{} then takes the top-100 candidates, shuffles them, stores the existing non-terminal tree nodes in a non-deterministic structure, and iteratively tries the merges, until one succeeds. For example, in Figure~\ref{fig:arvada-weaknesses}'s first bubbling step \arvada{} ranks 1,043 candidate 1- and 2-bubbles to merge the lime bubble ($t_1\to$~\term{L\textvisiblespace{}=\textvisiblespace{}n}).

\textbf{Accepting Rule Generalization Via Sampled Programs:}
\label{sec:check-bubble}
For both above cases (merging initial single nodes or a candidate bubble), \arvada{} accepts the merge if the black-box parser accepts up to 100~freshly generated programs. From all programs that exercise the proposed generalized grammar rule, \arvada{} samples these 100~programs (50~per merged side) uniformly.

\textbf{Final Step (Expand Terminals):}
At the end \arvada{} expands each terminal to a larger character class, so the grammar may accept tokens that were not in the seed programs. For example, \arvada{} tries to expand \texttt{$t_1\to$} \term{1} | \term{2} to all single digits, integers, or alphanumeric letters. \arvada{} then samples 10 strings, generates programs, and checks them via the parser. A grammar's terminals may thus differ across \arvada{} runs on the same seed inputs.

\subsubsection{Not Generalizing Recursively}

\label{sec:not-reapply}

\arvada{}'s second challenge is that it does not recursively reapply a rule generalization it just learned and thus on some runs needs additional expensive steps or gets stuck. For example, Figure~\ref{fig:arvada-weaknesses} shows 5/10 runs we observed \arvada{} pursue for the \examplesCombined{} input programs. 
As the second bubble (yellow) it grouped sibling nodes \term{(n+n)} under new bubble parent $t_2$ and merged $t_2$ with \term{n}'s (not shown) dummy parent into $t_2$.

While this bubble yields an appropriate generalized grammar rule 
($t_2\to$ \term{n} | \term{(}$t_2$\term{+}$t_2$\term{)}), \arvada{} does not recursively reapply this just learned rule to its parse trees---even though the sibling sequence \term{(}$t_2$\term{+}$t_2$\term{)} is now present in the right parse tree. Instead, \arvada{} re-ranks all bubbles (an expensive operation), picks and merges another bubble (orange), and thus gets stuck.

\subsubsection{Breaking Bracket-implied Nesting Structure}
\label{sec:break-nesting}

Many languages use matching 
round \term{(} \term{)}, 
square \term{[} \term{]}, 
and curly \term{\{} \term{\}} brackets 
to recursively nest concepts such as 
class and function definitions,
code blocks,
parameter lists,
array creation and access, and 
various other expressions.
\arvada{}'s third key challenge is that on some runs it prioritizes a bubble that conflicts with a parse tree's bracket-implied nesting structure and thus gets stuck.

For example, in some runs on the Figure~\ref{fig:arvada-weaknesses} input programs \examplesCombined{}, \arvada{} breaks the while language's numerical expression nesting, which is defined via matching round brackets. In these runs \arvada{} partially merges the bracket-wise unbalanced orange bubble (\term{)\textvisiblespace{};\textvisiblespace{}}$t_0$) with the implicit parent of the last closing bracket. \arvada{} then cannot further generalize the grammar. The resulting grammar is recursive. But for statement sequences (recursive applications of the semicolon) the it only allows very specific instantiations, i.e., each generated statement sequence must start with an assignment statement that contains at least one addition
(\term{L\textvisiblespace{}=\textvisiblespace{}(}$t_2$\term{+}$t_2$ \dots).

\section{Overview and Design}
\label{sec:treevada-overview}

We guided our design via feedback from running \arvada{} and our alternatives on \arvada{}'s seed programs for tinyc~\cite{arvada21ASE}. 
To prevent over-fitting we did not use feedback from any other programs or languages we used in the subsequent evaluation.

\subsection{Assumptions on Strings \& Brackets}

\toolName{}'s current heuristics build on two ``soft'' assumptions, i.e., that many languages 
(1) use \term{'} \term{"} quotes to wrap strings and
(2) use \term{(} \term{)} \term{[} \term{]} \term{\{} \term{\}} brackets for nesting. If a language (also) uses these characters for other purposes then \toolName{}'s F1 score may suffer.

\subsection{Pre-tokenizing Input Programs}
\label{sec:string_literals}

As many languages share basic tokenization (or lexing) rules (e.g., an identifier is separated by some non-identifier token from the following token), \toolName{} and \arvada{} first tokenize their input programs. Tokenizing both likely yields higher-quality grammars and is more efficient than on each run rediscovering common lexing rules via relatively expensive grammar inference.

Both approaches thus replace a sequence of elements of one of the four character classes (lower-case, uppercase, whitespace, or digits) with a new terminal. This leaves all brackets, punctuation, and other ``special'' characters as individual character terminals. For example, on the \examplesCombined{} input programs of Figure~\ref{fig:arvada-weaknesses}, \arvada{} and \toolName{} produce the same token sequence.

\subsubsection{Program Structure in String Literals}

While not part of the main three problems we focus on, we also notice that \arvada{} treats the contents of string literals as program structure. For example, during initial tokenization \arvada{} tokenizes the 7-character input fragment \term{"k\textvisiblespace{}:-)"} into 7~nodes. It may then bubble and merge some of these nodes and get stuck. We thus want to distinguish string literals from program elements.

As this is not the paper's main focus we use a simple heuristic that solves several scenarios that are common in many languages. Specifically, we notice that many languages wrap a string literal in single (\term{'}) or double (\term{"}) quotes. When it encounters either quote character \toolName{} thus groups all following characters until again encountering the same quote character. While this scheme cannot handle all cases (e.g., escaped quote characters), it tokenizes common simple cases correctly, e.g., \term{"k\textvisiblespace{}:-)"} into three tokens, one per double-quote character plus one for the string literal's content.

\subsection{Pre-structuring Parse Trees Along Brackets}
\label{sec:initial-tree}

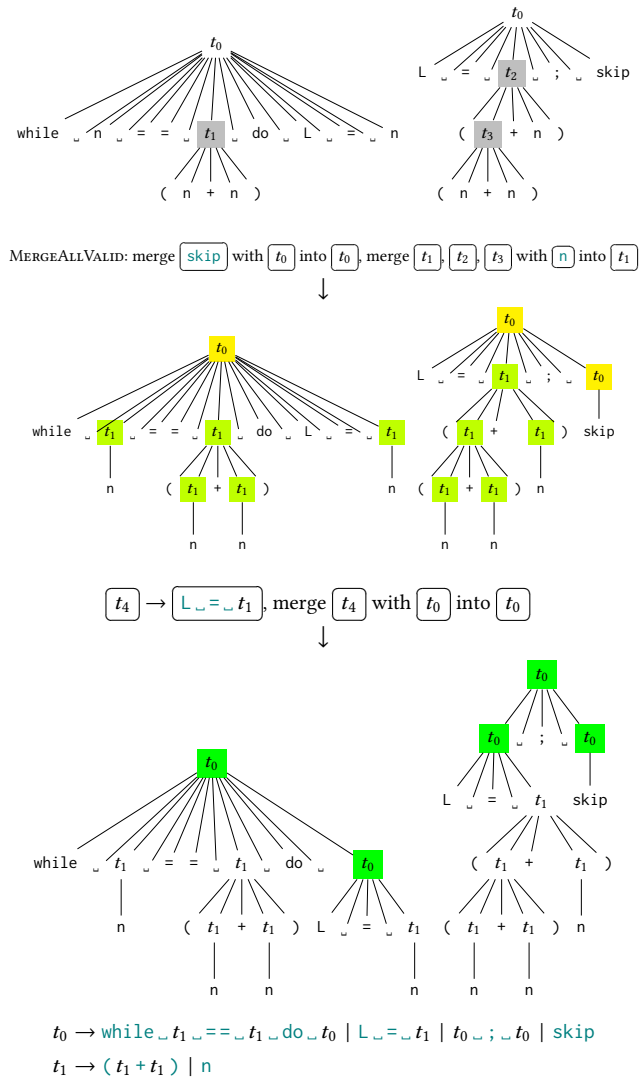
\begin{figure}[h!t]
    
\resizebox{\columnwidth}{!}{
	\begin{forest} [$t_0$, l sep={0.9cm}, for tree= {s sep={0.0mm}} 
	[\texttt{while}] 
	[\textvisiblespace{}]
	[\texttt{n}]
	[\textvisiblespace{}]
    [\texttt{=}][\texttt{=}]
    [\textvisiblespace{}]
    [$t_1$, fill=lightgray[\texttt{(}][\texttt{n}][\texttt{+}][\texttt{n}][\texttt{)}]]
    [\textvisiblespace{}]
	[\texttt{do}]
	[\textvisiblespace{}]
	[\texttt{L}]
	[\textvisiblespace{}]
	[\texttt{=}]
	[\textvisiblespace{}]
	[\texttt{n}]
	]	
	\end{forest} 
 \begin{forest}
	[$t_0$, l sep={0.3cm}, for tree= {s sep={0.0mm}} 
	    [\texttt{L}] 
	[\textvisiblespace{}]
	[\texttt{=}]
	[\textvisiblespace{}]
    [$t_2$,fill=lightgray[\texttt{(}][$t_3$,fill=lightgray[\texttt{(}][\texttt{n}][\texttt{+}][\texttt{n}][\texttt{)}]][\texttt{+}][\texttt{n}][\texttt{)}]]
	[\textvisiblespace{}]
	[\texttt{;}]
	[\textvisiblespace{}]
	[\texttt{skip}]
	  ]
\end{forest}
	}
 
\vspace{1 em}
\resizebox{\columnwidth}{!}{
\small \textsc{MergeAllValid:} merge \ovalbox{\term{skip}} with \ovalbox{$t_0$} into \ovalbox{$t_0$}, merge \ovalbox{$t_1$}, \ovalbox{$t_2$}, \ovalbox{$t_3$} with \ovalbox{\term{n}} into \ovalbox{$t_1$}
}
\vspace{-1em}
$$\downarrow$$
\resizebox{0.95\columnwidth}{!}{
	\begin{forest} [$t_0$,fill=yellow, l sep={0.9cm}, for tree= {s sep={0.0mm}} 
	[\texttt{while}] 
	[\textvisiblespace{}]
	[$t_1$,fill=lime[\texttt{n}]]
	[\textvisiblespace{}]
    [\texttt{=}][\texttt{=}]
    [\textvisiblespace{}]
    [$t_1$,fill=lime[\texttt{(}][$t_1$,fill=lime[\texttt{n}]][\texttt{+}][$t_1$,fill=lime[\texttt{n}]][\texttt{)}]]
    [\textvisiblespace{}]
	[\texttt{do}]
	[\textvisiblespace{}]
	[\texttt{L}]
	[\textvisiblespace{}]
	[\texttt{=}]
	[\textvisiblespace{}]
	[$t_1$,fill=lime[\texttt{n}]]
	]	
	\end{forest} 
 \begin{forest}
	[$t_0$,fill=yellow, l sep={0.3cm}, for tree= {s sep={0.0mm}} 
	    [\texttt{L}] 
	[\textvisiblespace{}]
	[\texttt{=}]
	[\textvisiblespace{}]
    [$t_1$,fill=lime[\texttt{(}][$t_1$,fill=lime[\texttt{(}][$t_1$,fill=lime[\texttt{n}]][\texttt{+}][$t_1$,fill=lime[\texttt{n}]][\texttt{)}]][\texttt{+}][$t_1$,fill=lime[\texttt{n}]][\texttt{)}]]
	[\textvisiblespace{}]
	[\texttt{;}]
	[\textvisiblespace{}]
	[$t_0$,fill=yellow[\texttt{skip}]]
	  ]
\end{forest}
	}
 
\vspace{1em}

{ \small  \ovalbox{$t_4$} $\to$  \ovalbox{\texttt{\term{L}\,\term{\textvisiblespace{}}\,\term{=}\,\term{\textvisiblespace{}}\,$t_1$}},
 merge \ovalbox{$t_4$} with \ovalbox{$t_0$} into \ovalbox{$t_0$}
 }
\vspace{-.3em}
$$\downarrow$$
\resizebox{0.95\columnwidth}{!}{
	\begin{forest} [$t_0$,fill=green, l sep={1cm}, for tree= {s sep={0.0mm}} 
	[\texttt{while}] 
	[\textvisiblespace{}]
	[$t_1$[\texttt{n}]]
	[\textvisiblespace{}]
    [\texttt{=}][\texttt{=}]
    [\textvisiblespace{}]
    [$t_1$[\texttt{(}][$t_1$[\texttt{n}]][\texttt{+}][$t_1$[\texttt{n}]][\texttt{)}]]
    [\textvisiblespace{}]
	[\texttt{do}]
	[\textvisiblespace{}]
	[$t_0$, fill=green[\texttt{L}]
	[\textvisiblespace{}]
	[\texttt{=}]
	[\textvisiblespace{}]
	[$t_1$[\texttt{n}]]]
	]	
	\end{forest} 
 \begin{forest}
	[$t_0$,fill=green, l sep={0.1cm}, for tree= {s sep={0.0mm}} 
	    [$t_0$,fill=green[\texttt{L}] 
	[\textvisiblespace{}]
	[\texttt{=}]
	[\textvisiblespace{}]
    [$t_1$[\texttt{(}][$t_1$[\texttt{(}][$t_1$[\texttt{n}]][\texttt{+}][$t_1$[\texttt{n}]][\texttt{)}]][\texttt{+}][$t_1$[\texttt{n}]][\texttt{)}]]]
	[\textvisiblespace{}]
	[\texttt{;}]
	[\textvisiblespace{}]
	[$t_0$,fill=green[\texttt{skip}]]
	  ]
\end{forest}
	}
\small
\begin{align*}
    t_0 \to~& \term{while}\, \term{\textvisiblespace{}}\, t_1\, \term{\textvisiblespace{}}\, \term{=}\, \term{=}\, \term{\textvisiblespace{}}\, t_1\, \term{\textvisiblespace{}}\, \term{do}\, \term{\textvisiblespace{}}\, t_0\, \, |\, \, \term{L}\, \term{\textvisiblespace{}}\, \term{=}\, \term{\textvisiblespace{}}\, t_1 \, \, |\, \, t_0\, \term{\textvisiblespace{}}\, \term{;}\, \term{\textvisiblespace{}}\, t_0\, \, |\, \, \term{skip}\\
    t_1 \to~& \term{(}\, t_1\, \term{+}\, t_1\, \term{)} \, \, |\, \, \term{n}
\end{align*}

\caption{Top to bottom: 
\toolName{}'s pre-structured bracket-implied trees for Figure~\ref{fig:arvada-weaknesses}'s \examplesCombined{} input programs with bracketed sequences (gray) bubbled, %
initial node-pair merges (yellow \& lime), and
1\textsuperscript{st} bubble via bubble-ranking (green);
the inferred grammar captures \examplesCombined{}'s Figure~\ref{fig:while} golden grammar rules.}
\label{fig:arvada-weaknesses-treevada}
\end{figure}

We observe that \arvada{}'s first bubble-ranking generalization step is its most expensive, as it may rank in $O(n^4)$ all pairs of all possible sibling token sequences of the input programs. 
Many such bubbles are likely illegal as they cross a round/square/curly bracket ``boundary'' and thus violate a nesting rule that is common among languages. This becomes clear on the extreme example program of $n = 2b+1$ tokens that starts with $b$ round opening brackets followed by \term{x} and $b$ round closing brackets. \arvada{}'s first bubble generalization step ranks $O(n^4)$ bubble pairs. Most such bubbles cross a nesting boundary and thus the parser likely rejects them.

The deeper-structured the parse-trees become via generalization steps, the cheaper each subsequent generalization step is. These later steps are cheaper not so much due to earlier grammar generalization but as a more-structured tree only permits shorter (and thus fewer) sibling token sequences. Our goal is thus to quickly convert parse trees from flat to richly structured, by essentially enforcing common nesting rules. In the above extreme nesting example, the nesting-implied parse tree consists of root node plus a single stack of $b$ layers of a single bracket-wrapped node, reducing $O(n^4)$ to a single $O(n^2)$ step, i.e., the upfront \textsc{MergeAllValid}.

Given the wide use of round/square/curly bracket-defined nesting, \toolName{} pre-structures parse trees heuristically---likely without significantly impairing the inferred grammar's quality. Specifically, \toolName{} makes one simple stack-based pass over each input program, initializing the stack (and the parse tree) with root $t_0$. \toolName{} adds each token to the parse tree as the child of the current top-of-stack node. When encountering an opening \term{(} \term{[} \term{\{} bracket, \toolName{} first pushes a new non-terminal onto the stack. When encountering a matching closing bracket \term{)} \term{]} \term{\}}, \toolName{} then pops the top element off the stack. When brackets no longer match \toolName{} reverts to a flat tree.

The subsequent attempt to merge all tree node pairs often merges some of these new rules with each other or other nodes. For example, the Figure~\ref{fig:arvada-weaknesses-treevada} run merges the bracket-implied $t_1$, $t_2$, and $t_3$ with the existing node \term{n}. If a pre-structured rule remains un-merged it does not generalize the grammar. As it just adds slightly to grammar verbosity, we do not remove such rules.

\arvada{}'s motivating \emph{while} example language uses brackets only lightly (i.e., only one rule in the Figure~\ref{fig:while} golden grammar contains brackets). But even then there are several cases (e.g., for the Figure~\ref{fig:arvada-weaknesses} input programs) where \toolName{} is faster, infers a better grammar, or does both. Figures~\ref{fig:arvada-weaknesses} and~\ref{fig:arvada-weaknesses-treevada} is an example of the latter.

\subsection{Removing Specialized Bubbling Heuristics}

As \toolName{} creates nesting structure upfront, there is less need for special cases and we therefore remove the following two rarely successful strategies \arvada{} uses.

\textbf{1-bracket Bubbles:}
Pre-structuring the parse trees ensures that each sibling node sequence contains at most two round, curly, or square brackets. While this prevents a bubble from crossing concept nesting, we observe that one bracket rarely generalizes the grammar correctly either. Not generating 1-bracket bubbles thus ensures that \toolName{} never considers a bracket-unbalanced bubble.

\textbf{Partial 1-char Node Merges:}
When \arvada{} cannot merge a given bubble with any interior node, it also tries merging the bubble with a subset of interior node instances that represent a terminal character, e.g., to merge the bubble with one  ``\term{)}'' instance but not others. As both \arvada{} and \toolName{} pre-tokenize their input programs, such special treatment of 1-char tokens rarely yields a successful merge and \toolName{} thus omits such partial merges.

\subsection{Deterministic Grammar Inference}
\label{sec:deterministic-treevada}

\toolName{} addresses \arvada{}'s sources of non-determinism. Several of these were easy to fix without significant performance degradation, just by switching the implementation to a deterministic data-structure. Specifically, \toolName{} orders the parse trees' unique node types by their shortest distance from any program root node (in ascending order) for two related operations. First, when trying to merge all node-pairs upfront and after exhausting bubbling. Second, when ordering merge-target nodes for a given bubble.

Similarly, when there are more than 50 candidate programs that exercise a candidate rule-merge on one of the two merged rules, \toolName{} makes \arvada{}'s program sampling strategy deterministic, by switching to deterministic data structures and always using the same random number generator seed value. Finally, \toolName{} makes \arvada{}'s terminal expansion deterministic, by fixing the random number generator's seed value to sample 10~programs from the larger character class. Following is a more complex case, where \toolName{} needed a new heuristic to compensate for \arvada{}'s benefits from arbitrary order and randomness.

\subsubsection{Depth- and Length-aware Bubble Ranking}
\label{sec:bubble-ranking}
To avoid \arvada{}'s non-deterministic shuffling of the top-100 ranked bubbles, \toolName{} builds on two observations. First, a longer bubble has a higher chance of being rejected because it tries to group together more nodes. A longer bubble, when accepted, also has a higher chance of getting grammar inference stuck. For example, in Figure~\ref{fig:arvada-weaknesses} \arvada{}'s last merge is a 5-node bubble $t_3\to$~\term{)\textvisiblespace{};\textvisiblespace{}}~$t_0$. The bubble's use of ``\term{)}'' prevents ``\term{)}'' from being used as the closing bracket in an otherwise possible subsequent bubbling of \term{(}$t_2$\term{+}$t_2$\term{)}.

Second, and more importantly, a more deeply nested bubble is more promising than a bubble closer to the root. The intuition is that a more deeply nested node sibling sequence is in a more specialized area of the input program that has more of its immediate surroundings already correctly structured via other rules. The likelihood of correctly generalizing such an already specialized area thus tends to be higher and, crucially, the impact of getting it wrong is lower---as it will only affect a relatively specialized program area. 

\toolName{} thus refines the bubble ranking, by adding two new criteria, bubble depth and bubble length. We have found that with these additions the bubble ranking is more reliable and does not require shuffling to get to promising candidate bubbles early on. The resulting ranking scheme ranks by context similarity first, then resolves ties via bubble depth (i.e., the bubble occurrences' minimum root-distance), further ties via bubble occurrence counts, and additional ties via bubble length.

\subsection{Applying Learned Rules Recursively}
\label{sec:solution_reapply}

After merging a bubble, \arvada{} does not recursively reapply the just learned grammar rule. For example, in Figure~\ref{fig:arvada-weaknesses} \arvada{} merged $t_2\to$~\term{(n+n)} with \term{n} into $t_2$, yielding rule \texttt{$t_2\to$ \term{(}$t_2$\term{+}$t_2$\term{)} | \term{n}}. While \arvada{} proceeds by re-ranking all bubbles, \toolName{} here instead directly tries to recursively reapply the just learned rule as renamed by the merge (i.e., $t_2\to$\term{(}$t_2$\term{+}$t_2$\term{)}). 

For the Figure~\ref{fig:arvada-weaknesses} scenario, \toolName{} would group \term{(}$t_2$\term{+}$t_2$\term{)} under a new $t_2$ node, yielding fewer direct children under the second parse tree's root node (\term{L\textvisiblespace=\textvisiblespace}$t_2$\term{\textvisiblespace;\textvisiblespace}$t_0$). This cheaply adds to the parse trees' structure we have just accepted as correct. In this case it would also prevent \arvada{} from its last bubble step that gets it stuck by breaking the parse tree's nesting structure.

\section{Evaluation}

Overall we would like to get a better understanding of how \toolName{} compares with the state-of-the-art approach \arvada{}, both on very small and slightly larger input programs. While the larger input programs may not yet be representative of how a user would want to apply these approaches on other languages, it at least gives us a glimpse of the scalability of the compared approaches. We thus seek to answer the following research questions.
\begin{description}
    \item[RQ0] Baseline: How does non-determinism affect \arvada{}? %
    \item[RQ1] Grammar quality: At similar runtime, does \toolName{} infer better grammars than \arvada{}?
    \item[RQ2] Runtime: When inferring grammars of similar quality, does \toolName{} have a lower runtime than \arvada{}?
    \item[RQ3] Readability: How compact are the inferred grammars?
    \item[RQ4] Ablation study: How do \toolName{}'s components influence its resource consumption and grammar quality?
\end{description}

To ease comparison, we run our experiments in \arvada{}'s Docker image\footnote{Accessed in January 2024: \url{https://github.com/neil-kulkarni/arvada}}. Specifically, from the image we reuse the \arvada{}, black-box parser, grammar-sampler, random program generator, and ANTLR4 parser generator~\cite{parr2014adaptive} binaries. From the image we also use the languages' existing 1k test programs. The following summarizes the metrics we reuse from \arvada{}'s work.

\textbf{Precision:}
From each \arvada{}-/\toolName{}-inferred grammar we sample 1k~programs and count how many of these 1k~programs the respective existing ``golden''
black-box parser accepts.

\textbf{Recall:}
We compile the \arvada{}-/\toolName{}-inferred grammar into a parser and count how many of the given 1k~(``golden'') test programs that parser accepts.

\textbf{F1 score:} 
As usual, the F1 score is the harmonic mean of precision and recall and ranges from 0 (zero precision or zero recall) to 1 (both perfect precision and perfect recall).

\textbf{Runtime:}
The main measure is the \arvada{}/\toolName{} runtime, which does not include computing precision or recall.

\textbf{Averages:}
We follow \arvada{}~\cite{arvada21ASE} in comparing a deterministic technique's result with the average of 10~non-deterministic runs. The latter estimates what a user may expect from running \arvada{} once. We also plot each of our \arvada{} and \toolName{} runs.

\subsection{RQ0: Timeouts vs. Precision Results}

To get a sense of non-determinism's effect on \arvada{} we first reproduce (``different team, same experimental setup'')\footnote{\url{https://www.acm.org/publications/policies/artifact-review-and-badging-current}} \arvada{}'s main results~\cite[Table~1]{arvada21ASE}---i.e., runtime and F1 score. We use \arvada{}'s Docker image (same \arvada{} configuration options, etc., including the same (``seed'') training input programs) and reran all languages from \arvada{}'s experiment as in the \arvada{} paper 10~times. Here we used a 24GB~RAM Ryzen-9 5900HX 3.30GHz~CPU laptop.

\begin{table}[h!]
\caption{\arvada{} study~\cite[Table~1]{arvada21ASE} (left); 
our rerun (middle), and \toolName{} (right) on the same input programs;
\arvada{} values are average over 10~runs;
f1~=~F1~score;
t~=~runtime;
$\pm$~=~standard deviation;
bold~=~unreliable precision \& F1 score. 
}

\begin{center}
\begin{small}
\begin{tabular}{l|rr|rr|rr}
    \toprule
& \multicolumn{2}{c|}{Earlier study~\cite{arvada21ASE}} 
& \multicolumn{2}{c|}{Rerun \arvada{}}
& \multicolumn{2}{c}{\toolName{}}\\
& \multicolumn{1}{c}{f1} 
& \multicolumn{1}{c|}{t[s]}
& \multicolumn{1}{c}{f1}
& \multicolumn{1}{c|}{t[s]}
& \multicolumn{1}{c}{f1}
& \multicolumn{1}{c}{t[s]} \\
    \midrule
    arith &1$\pm$.00& 3$\pm$0& \textbf{1$\pm$.00}& 1$\pm$0.1   \\
    fol &.91$\pm$.18& 372$\pm$36& .88$\pm$.19& 127$\pm$19 & \multicolumn{2}{c}{n/a} \\ 
    math. & .89$\pm$.08 & 65$\pm$6& .82$\pm$.13& 19$\pm$1.8   \\
    \midrule
    json & .97$\pm$.05 & 76$\pm$11 & .97$\pm$.04& 28$\pm$3.1 &.96&22 \\
    lisp  & .57$\pm$.21 & 16$\pm$4 & .58$\pm$.22& 6$\pm$1.2 &.71&7 \\
    turtle & 1$\pm$.00& 84$\pm$8& 1$\pm$.00 & 31$\pm$3.0  &1.0&25 \\
    while & .81$\pm$.14 & 54$\pm$5& .83$\pm$.12 & 20$\pm$1.6 &.38&16 \\
    xml & .96$\pm$.08 & 205$\pm$34&  .88$\pm$.30 & 78$\pm$8.2  &.09&88 \\
    \midrule
    curl & .68$\pm$.11 & 111$\pm$12&  .68$\pm$.10 & 60$\pm$6.9  &.63&102 \\
    tinyc & .81$\pm$.08 & 6.4k$\pm$1.2k& .69$\pm$.16 & 5k$\pm$1k &.81&1.5k \\
    nodejs & .29$\pm$.16 & 46k$\pm$22k&.39$\pm$.16 & 23k$\pm$6k &.56&4.5k \\   
    \bottomrule      
    \end{tabular}
\label{tab:rerun_Arvada_input_programs}
\end{small}
\end{center}
\end{table}

Table~\ref{tab:rerun_Arvada_input_programs} summarizes the results. First, in our rerun the runtime was consistently lower (likely due to the different machine). For the main grammar quality measure (F1~score), the average over all 11~languages was similar (80.8\% vs our rerun's 79.3\%). For individual languages the impact was larger. For example, tinyc's  average F1~score dropped from 81\% to 69\%. It may thus be misleading to compare \arvada{} performance across languages. For example, while in the earlier study \arvada{} produced much better grammars for tinyc than for curl (81\% vs 68\%), this difference all but disappeared in our rerun (69\% vs 68\%).

From the source code we learnt that when calling a black-box parser \arvada{}'s precision calculation enforces a 10s~timeout. While the programs are just a few dozen tokens, arith's ``golden'' parser timed out for many programs sampled from the \arvada{}-inferred grammars, with parse time growing quickly with program size. On one example arith run 76/1k programs timed out. \arvada{}'s metric tool unfortunately treats a timeout as if the parser accepted the program, which very likely corrupts precision (and thus F1 score). As we could not easily solve the problem (e.g., by increasing the timeout by 10$\times$), we exclude arith from tool comparisons.

Excluding a language for unreliable measurement is not meant to avoid running \toolName{}. On the Table~\ref{tab:rerun_Arvada_input_programs} arith seeds \toolName{} ``scores'' 100\% precision via \arvada{}'s metric tool. Due to the metric tool's parser timeout treatment, this result is equally unreliable.

\subsection{Experimental Setup for RQ1 to RQ4}

While the \arvada{} work carefully constructed minimal input programs that cover all rules of a given golden grammar, we aim to emulate a more realistic scenario where the user does not have a golden grammar and thus cannot construct a minimal set of minimal input programs. Hence we rely on the \arvada{} Docker image's 1k test programs. These programs do not guarantee to cover all golden grammar rules. For the first 5/8 languages, we randomly pick seed inputs from this pool of 1k programs (which may slightly inflate recall but still allows comparing \toolName{} with \arvada{}). 

\begin{table}[ht!]
    \caption{
    Input programs $S$ with their avg/max \toolName{} pre-tokenized tokens;
    \#~=~nr. programs same for R1 and R2;
    Q~=~programs containing a \term{"} or \term{'};
    c~=~tinyc;
    js~=~nodejs.
    }
    \begin{center}
\resizebox{\columnwidth}{!}{
\begin{tabular}{l|crr|r|rrr|rrr}
    \toprule
    & \multicolumn{3}{c|}{$S$\textsubscript{H/R0\cite{arvada21ASE}}} 
    & \multicolumn{1}{c}{}
    & \multicolumn{3}{c}{$S$\textsubscript{R1/R5}}
    & \multicolumn{3}{c}{$S$\textsubscript{R2}} \\
    & \multicolumn{1}{c}{\#}
    & \multicolumn{1}{c}{avg}
    & \multicolumn{1}{c|}{m}
    & \multicolumn{1}{c|}{\#}
    & \multicolumn{1}{c}{avg}
    & \multicolumn{1}{c}{m}
    & \multicolumn{1}{c|}{Q}
    & \multicolumn{1}{c}{avg}
    & \multicolumn{1}{c}{m}
    & \multicolumn{1}{c}{Q}
    \\
\midrule
{json}      & 71    &3.0& 13&30    & 8.2     & 72 &20 & 5.4&    25 &15\\
{lisp}      & 26    &2.3& 7&30   & 76.7    & 418 &0 &23.8&   120 &0\\ 
{turtle}    & 33    &7.0& 13&35   & 30.2     & 91  &0 &19.2&    63 &0\\
{while}     & 10    &7.8& 12&30    &110.7   &534  &0 &135.1&    1298 &0\\
{xml}       & 40    &11.4& 20&20    & 24.4    & 62 &14 &25.4&    75 &13\\
\midrule
{curl}      & 25    &13.7& 25&25    & 15.0     & 26 &1 &14.6&   29 &0\\
{c}     & 25    &69.1& 207&25    & 81.5   & 218  &0 &72.7&   213  &0\\
{js}    & 25    &49.0& 146&15    &60.0    &116  &4 &65.9&    244 &5\\
\midrule
{c-500}     &      & n/a&       &10     &420.3   &483  &0 & \multicolumn{3}{c}{n/a} \\
{js-500}    &      &n/a&       &5     &307.0   &385 &2 &  \multicolumn{3}{c}{n/a}  \\
\bottomrule      
\end{tabular}
}
\label{tab:input-program-stats}
\end{center}
\end{table}

As in the \arvada{} work, for the next 3/8 languages (curl, tinyc, and nodejs) we do not have a golden grammar. We thus use the same 3rd-party random program generators the \arvada{} work used (with their default settings) to create new seed programs. For all 8~languages the new input sets $S$ may thus not cover all golden grammar rules. We call this random input set $R1$.

Table~\ref{tab:input-program-stats} compares the $R1$ input programs with the ones used in the \arvada{} work---the first 5 languages had handpicked seeds (``H'') and the next 3 used 3rd-party generators (``R0''). We focus on the token counts via \toolName{}'s tokenization scheme, as it only differs in how it treats \term{"} and \term{'}, which only a few input programs contain. Compared to the \arvada{} study, the average token count tends to be larger---for most of the first 5 languages by an order of magnitude. Especially the largest programs are significantly larger. 

To explore larger input programs we generate another set $R5$ using the same generator used for tinyc~\cite{gopinath20mining} (``tinyc-500'') and nodejs~\cite{padhye2019semantic} (``nodejs-500''). Here we skip programs under 200 characters long. Table~\ref{tab:input-program-stats} shows tinyc-500 and nodejs-500 programs are on average 5$\times$ larger than tinyc and nodejs programs by token count. R1/R5 seed (and test) programs either both did (json, xml, curl, js, js-500) or did not have some programs with quotes. All R1/R5 seed (and test) programs had brackets, except for xml. Curl had one seed program with unmatched brackets.

Here we run each experiment on an EPYC Milan 7763 64-core~CPU machine in TACC's Lonestar6 cluster\footnote{Accessed in January 2024: \url{https://tacc.utexas.edu/systems/lonestar6}}, which does not support Docker. We thus recreated the Docker image's setup as closely as possible (same oracle binaries, etc.). After removing arith we removed the parser timeouts. First, we removed a 3s parser timeout \arvada{} used for its grammar inference. This should improve the quality of \arvada{}-inferred grammars as it no longer has to interpret a parser timeout as ``parsed ok''. 

After removing the 3s parser timeouts, sampled programs of fol and math also got stuck (for at least 30 minutes each) in their ``golden'' parser (due to poorly written grammars). To protect precision calculation's integrity we also removed fol and math. For precision calculation we could then remove its 10s~parser timeout, which allowed us to just use the first 1k  programs sampled from an \arvada{}-/\toolName{}-inferred grammar (\arvada{}'s evaluation~\cite{arvada21ASE} silently discarded any sampled program over 300~characters).

We started all experiments with 32GB~RAM. As \arvada{}'s grammar inference ran out of memory on lisp, nodejs, and nodejs-500, for these three experiments we then used an otherwise identically configured 256GB~RAM machine.

\subsection{RQ1, RQ2: Precision, Recall, F1, Resources}

\begin{table*}[!htb]
\caption{Average \arvada{} \& \toolName{} results over 10 runs;
r~=~recall;
p~=~precision;
f1~=~F1~score;
t~=~runtime;
t\textsubscript{O}~=~oracle time;
q~=~queries sent to oracle;
m~=peak memory usage;
$\pm$~=~standard deviation;
bold~=~\toolName{} $\geq 2\times$ better in main metrics.
}
\begin{center}
\begin{small}
\begin{tabular}{l|rrrr|rrr|rrrr|rrr}
\toprule
& \multicolumn{7}{c|}{\arvada{} on randomly selected seeds: R1, R5 (R5 = c-500 \& js-500)} &  \multicolumn{7}{c}{\toolName{} on same set of random seeds, m in GB} \\
Name 
& \multicolumn{1}{c}{r}
& \multicolumn{1}{c}{p}
& \multicolumn{1}{c}{f1}
& \multicolumn{1}{c}{t[ks]}
& \multicolumn{1}{c}{t\textsubscript{O}[ks]} 
& \multicolumn{1}{c}{q[k]}
& \multicolumn{1}{c|}{m[GB]}
& \multicolumn{1}{c}{r}
& \multicolumn{1}{c}{p}
& \multicolumn{1}{c}{f1}
& \multicolumn{1}{c}{t[ks]}
& \multicolumn{1}{c}{t\textsubscript{O}[ks]} 
& \multicolumn{1}{c}{q[k]}
& \multicolumn{1}{c}{m}\\
\midrule
json	&.97	$\pm$.02	&.68	$\pm$.13	&.79	$\pm$.09	&.13	$\pm$.0	&.08	$\pm$.0	&18.0	$\pm$4.2	&.05	$\pm$.0	&.90	&.96	&.93	&\textbf{.03	$\pm$.0}	&.03	$\pm$.0	&9.0	&.03	\\
lisp	&.53	$\pm$.38	&.93	$\pm$.13	&.57	$\pm$.27	&6.1	$\pm$1.5	&.44	$\pm$.4	&35.8	$\pm$30.9	&25.7	$\pm$20.2	&1.0	&.98	&.99	&\textbf{.70	$\pm$.0}	&.39	$\pm$.0	&58.3	&\textbf{.06}	\\
turtle	&1.0	$\pm$.01	&.95	$\pm$.08	&.97	$\pm$.05	&.56	$\pm$.1	&.17	$\pm$.1	&29.1	$\pm$8.3	&.06	$\pm$.0	&1.0	&.94	&.97	&\textbf{.12	$\pm$.0}	&.08	$\pm$.0	&16.7	&.04	\\
while	&1.0	$\pm$.00	&.99	$\pm$.03	&1.0	$\pm$.01	&7.8	$\pm$.9	&1.1	$\pm$.2	&53.2	$\pm$9.6	&.76	$\pm$.2	&1.0	&1.0	&1.0	&\textbf{2.1	$\pm$.0}	&.19	$\pm$.0	&34.2	&\textbf{.13}	\\
xml	&1.0	$\pm$.00	&.70	$\pm$.37	&.76	$\pm$.30	&.30	$\pm$.0	&.12	$\pm$.0	&25.2	$\pm$3.8	&.05	$\pm$.0	&1.0	&1.0	&1.0	&.16	$\pm$.0	&.05	$\pm$.0	&12.5	&.08	\\
\midrule
curl	&.90	$\pm$.09	&.71	$\pm$.14	&.78	$\pm$.09	&.15	$\pm$.0	&.12	$\pm$.0	&20.8	$\pm$2.1	&.05	$\pm$.0	&.60	&.89	&.72	&.12	$\pm$.0	&.09	$\pm$.0	&20.3	&.05	\\
tinyc	&.76	$\pm$.20	&.54	$\pm$.19	&.62	$\pm$.18	&7.4	$\pm$1.0	&1.9	$\pm$.6	&131.5	$\pm$27.2	&.77	$\pm$.1	&.97	&.86	&.91	&\textbf{3.1	$\pm$.0}	&.56	$\pm$.0	&164.5	&\textbf{.22}	\\
nodejs	&.06	$\pm$.07	&.25	$\pm$.09	&.08	$\pm$.09	&8.7	$\pm$3.3	&5.1	$\pm$2.3	&130.5	$\pm$69.2	&3.1	$\pm$1.7	&\textbf{.47}	&\textbf{.82}	&\textbf{.59}	&5.1	$\pm$.0	&4.8	$\pm$.0	&169.2	&\textbf{.07}	\\
\midrule
c-500	&.55	$\pm$.20	&.81	$\pm$.08	&.63	$\pm$.16	&29.6	$\pm$4.7	&9.4	$\pm$3.1	&247.8	$\pm$82.7	&2.4	$\pm$1.1	&.53	&.93	&.67	&\textbf{6.8	$\pm$.2	}&.85	$\pm$.0	&175.5	&\textbf{.40}	\\
js-500	&.12	$\pm$.15	&.26	$\pm$.19	&.12	$\pm$.15	&23.0	$\pm$5.0	&13.0	$\pm$3.6	&218.9	$\pm$56.7	&8.9	$\pm$2.7	&\textbf{.37}	&\textbf{.62}	&\textbf{.46}	&16.7	$\pm$.3	&14.4	$\pm$.3	&480.4	&\textbf{.35}	\\
\bottomrule
\end{tabular}
\label{tab:perf-results}
\end{small}
\end{center}
\end{table*}

Table~\ref{tab:perf-results} shows the main evaluation results (on R1 and R5). Across all 10~experiments, \toolName{} on average both produces better grammars and is faster than \arvada{}, i.e., \toolName{} has a 9.3\% higher recall, a 22.1\% higher precision, a 19.5\% higher F1~score, and a 2.4$\times$ speedup over \arvada{}. In 9/10 experiments \toolName{} inferred a grammar of the same or higher quality than the average \arvada{} run (i.e., \toolName{}'s F1~score was at least as high as \arvada{}'s). 

\begin{figure*}
\begin{tikzpicture}[font=\sffamily]
\begin{groupplot}
    [
        group style=
            {
            columns=10,
            xlabels at=edge bottom,
            y descriptions at=edge left, %
            horizontal sep=0.01\textwidth, %
            group name=plots
            },
        ybar,
        /pgf/bar width=3pt, %
        scale only axis,
        enlarge x limits = 0.70,
        width=0.9\textwidth/11, %
        height=\textwidth/7, %
        ylabel=\% of \arvada{}'s total runtime,
        ylabel style={
          font=\sffamily\tiny,
          inner ysep=0pt
          },
        xticklabel style={font=\sffamily\tiny}, %
        tickpos=left,
        legend image code/.code={%
             \draw[#1] (0cm,-0.1cm) rectangle (0.3cm,0.1cm);
        },
        yticklabel style={font=\sffamily\scriptsize},
        y tick scale label style={at={(yticklabel* cs:1.03, 2em)}}, %
        title style={anchor=base}, %
        axis y line*=left,
        axis x line*=bottom,
        ymin=0,
        xtick=data,
        legend style={at={(-2.8,1.4)},font=\sffamily\scriptsize,
legend columns=-1},
        symbolic x coords={A,T}
    ]

\nextgroupplot[title=\scriptsize json]
\addplot+ [pattern=dots,
            error bars/.cd,
                y dir=both,
                y explicit,
        ] 
            coordinates {
            (A,.06) +- (0,0.01)
            (T,0.01) 
            };
          \addplot+ [pattern=north east lines,
            error bars/.cd,
                y dir=both,
                y explicit,
        ]
            coordinates {
            (A,.13) +- (0,.05)
            (T,0.03) 
            };
            \addplot+ [pattern=north west lines,
            error bars/.cd,
                y dir=both,
                y explicit,
        ]
            coordinates {
            (A,.62) +- (0,.13)
            (T,.21)
            };
\nextgroupplot[title=\scriptsize lisp, hide y axis]
\addplot+ [pattern=dots,
            error bars/.cd,
                y dir=both,
                y explicit,
        ] 
            coordinates {
            (A,.73) +- (0,.35)
            (T,0.0) 
            };
          \addplot+ [pattern=north east lines,
            error bars/.cd,
                y dir=both,
                y explicit,
        ]
            coordinates {
            (A,.12) +- (0,.02)
            (T,0.02) 
            };
            \addplot+ [pattern=north west lines,
            error bars/.cd,
                y dir=both,
                y explicit,
        ]
            coordinates {
            (A,.07) +- (0,.07)
            (T,.07) 
            };

\nextgroupplot[title=\scriptsize turtle, hide y axis]
\addplot+ [pattern=dots,
            error bars/.cd,
                y dir=both,
                y explicit,
        ] 
            coordinates {
            (A,.07) +- (0,.03)
            (T,0.01) 
            };
          \addplot+ [pattern=north east lines,
            error bars/.cd,
                y dir=both,
                y explicit,
        ]
            coordinates {
            (A,.36) +- (0,.08)
            (T,.05)
            };
            \addplot+ [pattern=north west lines,
            error bars/.cd,
                y dir=both,
                y explicit,
        ]
            coordinates {
            (A,.30) +- (0,.09)
            (T,.13)
            };

\nextgroupplot[title=\scriptsize while, hide y axis]
\addplot+ [pattern=dots,
            error bars/.cd,
                y dir=both,
                y explicit,
        ] 
            coordinates {
            (A,.40) +- (0,.09)
            (T,0.02)
            };
          \addplot+ [pattern=north east lines,
            error bars/.cd,
                y dir=both,
                y explicit,
        ]
            coordinates {
            (A,.25) +- (0,.07)
            (T,.13)
            };
            \addplot+ [style={fill=green},pattern=north west lines,
            error bars/.cd,
                y dir=both,
                y explicit,
        ]
            coordinates {
            (A,.14) +- (0,.02)
            (T,.02)
            };

\nextgroupplot[title=\scriptsize xml, hide y axis]
\addplot+ [pattern=dots,
            error bars/.cd,
                y dir=both,
                y explicit,
        ] 
            coordinates {
            (A,.07) +- (0,0.02)
            (T,.09)
            };
          \addplot+ [pattern=north east lines,
            error bars/.cd,
                y dir=both,
                y explicit,
        ]
            coordinates {
            (A,.23) +- (0,.05)
            (T,.15) 
            };
            \addplot+ [pattern=north west lines,
            error bars/.cd,
                y dir=both,
                y explicit,
        ]
            coordinates {
            (A,.40) +- (0,.06)
            (T,.18) 
            };
\nextgroupplot[title=\scriptsize curl, hide y axis]
\addplot+ [pattern=dots,
            error bars/.cd,
                y dir=both,
                y explicit,
        ] 
            coordinates {
            (A,.04) +- (0,0.01)
            (T,0.04)
            };
          \addplot+ [pattern=north east lines,
            error bars/.cd,
                y dir=both,
                y explicit,
        ]
            coordinates {
            (A,.07) +- (0,0.01)
            (T,.05) 
            };
            \addplot+ [pattern=north west lines,
            error bars/.cd,
                y dir=both,
                y explicit,
        ]
            coordinates {
            (A,.83) +- (0,.07)
            (T,.62)
            };
\nextgroupplot[title=\scriptsize tinyc, hide y axis]
\addplot+ [pattern=dots,
            error bars/.cd,
                y dir=both,
                y explicit,
        ] 
            coordinates {
            (A,.41) +- (0,.04)
            (T,.04) 
            };
          \addplot+ [pattern=north east lines,
            error bars/.cd,
                y dir=both,
                y explicit,
        ]
            coordinates {
            (A,.16) +- (0,.03)
            (T,.15)
            };
            \addplot+ [pattern=north west lines,
            error bars/.cd,
                y dir=both,
                y explicit,
        ]
            coordinates {
            (A,.26) +- (0,.08)
            (T,.08)
            };

\nextgroupplot[title=\scriptsize nodejs, hide y axis]
\addplot+ [pattern=dots,
            error bars/.cd,
                y dir=both,
                y explicit,
        ] 
            coordinates {
            (A,.34) +- (0,.10)
            (T,0.00) 
            };
          \addplot+ [pattern=north east lines,
            error bars/.cd,
                y dir=both,
                y explicit,
        ]
            coordinates {
            (A,0.04) +- (0,0.02)
            (T,0.02)
            };
            \addplot+ [pattern=north west lines,
            error bars/.cd,
                y dir=both,
                y explicit,
        ]
            coordinates {
            (A,.58) +- (0,.26)
            (T,.55)
            };

\nextgroupplot[title=\scriptsize c-500, hide y axis]
\addplot+ [pattern=dots,
            error bars/.cd,
                y dir=both,
                y explicit,
        ] 
            coordinates {
            (A,.38) +- (0,.07)
            (T,.04) 
            };
          \addplot+ [pattern=north east lines,
            error bars/.cd,
                y dir=both,
                y explicit,
        ]
            coordinates {
            (A,.16) +- (0,.04)
            (T,.09) 
            };
            \addplot+ [pattern=north west lines,
            error bars/.cd,
                y dir=both,
                y explicit,
        ]
            coordinates {
            (A,.32) +- (0,.10)
            (T,.03) 
            };
            
\nextgroupplot[title=\scriptsize js-500, hide y axis]
\addplot+ [pattern=dots,
            error bars/.cd,
                y dir=both,
                y explicit,
        ] 
            coordinates {
            (A,.39) +- (0,.06)
            (T,0.00) 
            };
          \addplot+ [pattern=north east lines,
            error bars/.cd,
                y dir=both,
                y explicit,
        ]
            coordinates {
            (A,0.03) +- (0,0.01)
            (T,0.04) 
            };
            \addplot+ [pattern=north west lines,
            error bars/.cd,
                y dir=both,
                y explicit,
        ]
            coordinates {
            (A,.56) +- (0,.16)
            (T,0.63)
            };
\legend{Bubble Ranking, String sampling, Oracle calls}
\end{groupplot}
\end{tikzpicture}
\caption{Average (and standard deviation) of time spent on ranking bubbles, sampling strings, and in black-box parser. Each value is normalized by dividing by \arvada{}'s average total runtime for that language (R1, R5 seed);
A~=~\arvada{};
T~=~\toolName{}.
}
\label{fig:bar_chart}
\end{figure*}
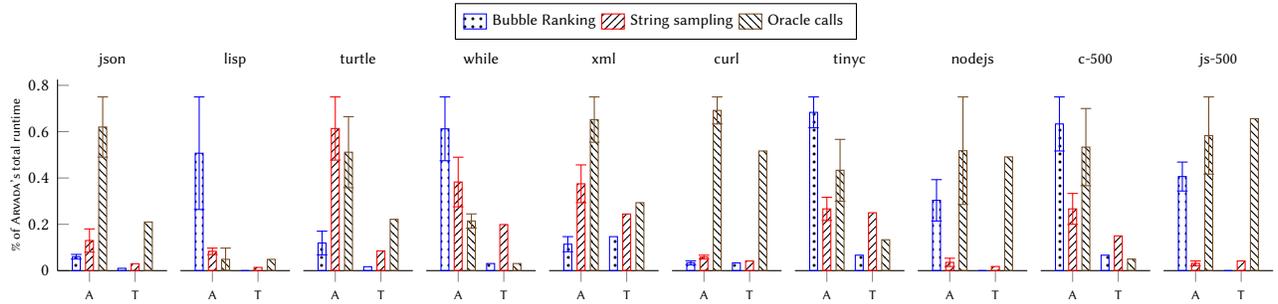

The outlier is curl, where \toolName{}'s 72\% F1 score is slightly below \arvada{}'s 78\%. curl has brackets but does not use them for nesting. \toolName{}'s attempts to pre-structure the input programs' parse trees thus either fail quickly during the initial pass over the input programs or get \toolName{} stuck with a sub-optimal grammar.

Compared to our reruns of the earlier study (Table~\ref{tab:rerun_Arvada_input_programs}, middle), \arvada{} increased its F1 score on 2/8~languages---i.e., for curl from 68 to 78\% and while from 83 to 100\%, likely due to non-determinism and differences in input programs. On the other hand, switching from hand-picked minimal programs to randomly selected input programs here may have contributed to lowering \arvada{}'s F1 score in 4/8~other languages---json from 97 to 79\%, xml from 88 to 76\%, tinyc from 69 to 62\%, and nodejs from 39 to 12\%.

At the same time, in all 10 experiments \toolName{} was faster than the average \arvada{} runtime. For 6/10 experiments \toolName{} was at least twice as fast as \arvada{}'s average run. On the larger input programs tinyc-500 and nodejs-500 (which have about double the total number of input tokens as tinyc and nodejs), \toolName{} remains faster than the average \arvada{} run and achieves higher F1~scores, i.e., 63 vs. 67\%  on tinyc-500 and 12 vs. 46\% on nodejs-500.

For scalability it is further interesting to consider the group of 7/10 experiments with the highest total token counts in their input programs---i.e., lisp, turtle, while, tinyc, nodejs, tinyc-500, and nodejs-500. In these 7~experiments \toolName{} has either much better grammar quality at similar runtime, much better runtime at similar grammar quality, or both much better grammar quality and runtime. For example, on the experiment with the highest total token count (tinyc-500 with some total 4.2k tokens), \toolName{}'s F1 score is 67\% vs \arvada{}'s 63\% while using less than a quarter of the time.
To better understand the differences between \arvada{} and \toolName{}, we further compare the following metrics.

\begin{table*}[!htb]
\caption{Grammar size and parse performance on R1 \& R5: Averages for grammars inferred in 10 Table~\ref{tab:perf-results} runs;
NT/T~=~unique {(non-)} terminals;
A~=~rules (alternatives);
l(A)~=~avg. rule length;
S~=~sum of rule lengths;
t\textsubscript{P}~=~time to parse 1k~``golden'' test programs;
m\textsubscript{P}~=~peak memory while parsing;
$\pm$~=~standard deviation (\toolName{} all zero);
bold~=~\toolName{} $\geq 2 \times$ better.
}
    \begin{center}
    \begin{small}
\begin{tabular}{l|crrrr|rr|rrrrr|rr}
    \toprule
    & \multicolumn{7}{c|}{\arvada{}} 
    & \multicolumn{7}{c}{\toolName{}, m\textsubscript{P} in GB} \\
    & \multicolumn{1}{c}{NT} 
    & \multicolumn{1}{c}{A} 
    & \multicolumn{1}{c}{l(A)}
    & \multicolumn{1}{c}{S}
    & \multicolumn{1}{c}{T}
    & \multicolumn{1}{c}{t\textsubscript{P}[ks]}
    & \multicolumn{1}{c|}{m\textsubscript{P}[GB]}
    & \multicolumn{1}{c}{NT} 
    & \multicolumn{1}{c}{A} 
    & \multicolumn{1}{c}{l(A)}
    & \multicolumn{1}{c}{S}
    & \multicolumn{1}{c}{T}
    & \multicolumn{1}{c}{t\textsubscript{P}[ks]}
    & \multicolumn{1}{c}{m\textsubscript{P}}\\ 
    \midrule
json	&30	$\pm$4	&153	$\pm$11	&1.4	$\pm$.0	&208	$\pm$19.8	&95	$\pm$1	&.02	$\pm$.0	&.03	$\pm$.0	&17	&178	&1.2	&208	&82	&\textbf{.01}	&.03	\\
lisp	&18	$\pm$10	&91	$\pm$27	&14.2	$\pm$9.9	&1040	$\pm$639.2	&40	$\pm$0	&.14	$\pm$.2	&.04	$\pm$.0	&18	&95	&1.8	&\textbf{171}	&40	&.08	&.04	\\
turtle	&28	$\pm$6	&123	$\pm$17	&1.6	$\pm$.2	&197	$\pm$43.9	&67	$\pm$0	&.03	$\pm$.0	&.04	$\pm$.0	&15	&95	&1.4	&131	&67	&.02	&.03	\\
while	&33	$\pm$5	&85	$\pm$13	&2.3	$\pm$.1	&191	$\pm$26.6	&18	$\pm$0	&.30	$\pm$.2	&.22	$\pm$.1	&24	&55	&2.3	&127	&18	&\textbf{.09}	&\textbf{.08}	\\
xml	&20	$\pm$3	&95	$\pm$7	&1.6	$\pm$.1	&149	$\pm$13.8	&58	$\pm$0	&.02	$\pm$.0	&.03	$\pm$.0	&9	&78	&2.0	&157	&58	&.02	&.03	\\
\midrule
curl	&22	$\pm$2	&174	$\pm$28	&1.4	$\pm$.1	&237	$\pm$29.1	&91	$\pm$14	&.80	$\pm$.5	&.07	$\pm$.0	&16	&137	&1.4	&189	&82	&\textbf{.16}	&.04	\\
tinyc	&57	$\pm$7	&229	$\pm$22	&1.9	$\pm$.2	&436	$\pm$31.9	&52	$\pm$3	&.70	$\pm$.7	&.08	$\pm$.1	&30	&149	&1.8	&264	&50	&\textbf{.34}	&.05	\\
nodejs	&46	$\pm$9	&273	$\pm$43	&2.6	$\pm$.6	&699	$\pm$66.4	&95	$\pm$3	&.16	$\pm$.2	&.04	$\pm$.0	&36	&196	&1.8	&351	&95	&.12	&.04	\\
\midrule
c-500	&63	$\pm$7	&246	$\pm$30	&2.2	$\pm$.6	&514	$\pm$75.7	&51	$\pm$2	&.60	$\pm$.3	&.09	$\pm$.0	&41	&187	&1.8	&345	&50	&\textbf{.26}	&.05	\\
js-500	&53	$\pm$6	&319	$\pm$38	&3.0	$\pm$.4	&935	$\pm$47.8	&95	$\pm$1	&10.6	$\pm$10.4	&9.9	$\pm$10.0	&45	&319	&1.9	&596	&102	&\textbf{.43}	&\textbf{.10}	\\
\bottomrule      
    \end{tabular}
\label{tab:grammar-analysis}
\end{small}
\end{center}
\end{table*}

\begin{figure*}[!h]
    \centering
    \includegraphics[width=\linewidth,trim={.1in 3.1in .1in 3.1in},clip]{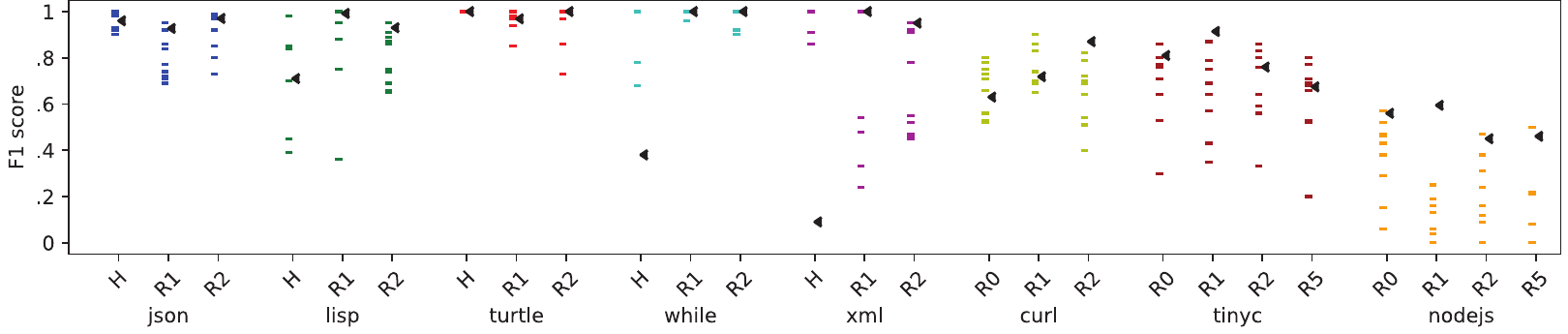}
    \caption{F1 score of 10~\arvada{}~(-) and \toolName{}~($\blacktriangleleft$) runs on hand-picked (H~\cite{arvada21ASE}) and random seeds~(R0~\cite{arvada21ASE}, R1, R2, R5).
    }
    \label{fig:plot-all-runs}
\end{figure*}

\textbf{Bubble Ranking Time:} Both \arvada{} and \toolName{} rank their potential grammar generalization steps (aka bubbles). The main difference is that \toolName{} tries to omit from this ranking bubbles that may violate common bracket-defined nesting rules.
   
\textbf{String Sampling Time:} Except for the use of non-determinism, \arvada{} and \toolName{} use the same scheme for sampling programs that exercise a proposed grammar rule merge.

\textbf{Oracle Calls \& Time:} \arvada{} and \toolName{} call an external parser during grammar inference in the same way.

\textbf{Memory Use:} We measure the peak memory use during grammar inference via Linux's \emph{time} command.

\textbf{Results:}
\toolName{} consistently uses the same or less memory than \arvada{}. In 3/10 experiments this difference is one (nodejs, nodejs-500) or even two (lisp) orders of magnitude. Figure~\ref{fig:bar_chart} reinforces that the source of this difference is the large difference in time spent on bubble ranking. Given that in these three experiments \toolName{} also yields significantly higher F1 scores makes clear that much of \arvada{}'s bubble ranking is counter-productive as it prioritizes bubbles that ultimately get the grammar inference stuck.

Bubble ranking is the dominating time expenditure for 4/10 of the \arvada{} experiments, but in none of the \toolName{} experiments. Instead the external parser dominates \toolName{}'s overall runtime in 7/10 experiments and in 6/7 of these cases by a large margin. In the remaining 3/10 cases \toolName{}'s bottleneck is string sampling.

\subsection{RQ3: Grammar Readability}

Neither \arvada{} nor \toolName{} attempt to simplify their inferred grammars. They just export the state of the grammar when they cannot find any additional grammar generalization steps. Since there are use-cases involving human consumption such as program understanding, it is still interesting to determine if higher grammar quality comes at the expense of larger grammars. 

A related question is if a larger grammar for a given language is maybe structured more efficiently for parsing, i.e., in parser runtime and memory consumption. To explore these two related questions we thus measure the following two metrics.

\textbf{Grammar Size:}
We count a grammar's unique non-terminals, unique terminals, number of rules (i.e., rule alternatives), and each rule's length (i.e., the length of a rule's right-hand-side sequence of terminals and non-terminals). The grammar's size is then the sum its rule lengths.

\textbf{Parse Time \& Memory:}
We measure the total time required and peak memory used to parse the 1k~``golden'' test programs using a parser generated from an \arvada{}-/\toolName{}-inferred grammar.

\textbf{Results:}
Table~\ref{tab:grammar-analysis} gives an overview of grammar size and parse performance. Despite covering more of the golden grammars (higher recall) and having higher F1 scores, \toolName{}'s grammars are smaller for 8/10 languages. The biggest difference is lisp where \toolName{}'s grammar is less than one sixth the size of the average \arvada{} grammar. For json, \toolName{}'s grammar size equals \arvada{}'s average grammar size. The only outlier is xml (157 vs. 149)---where \toolName{} has a significantly higher F1 score.

\arvada{}'s larger grammars do not improve parsing performance, as there is no experiment in which \arvada{}'s average parse time or memory use is lower than \toolName{}'s. On the contrary, for 6/10 experiments, \toolName{}'s parse time is less than half of \arvada{}'s.

\subsection{Performance Variance Across Seed Sets: R2}

To compare performance across seeds we generate a fresh round of random input program sets (``R2'') in the same style as for R1. Table~\ref{tab:input-program-stats} shows R2's size metrics. In R2 all seed sets contain some brackets, except for xml and curl.

\begin{figure}[!h]
    \centering
    \includegraphics[width=\columnwidth,trim={2.9in 3in 3in 3in},clip]{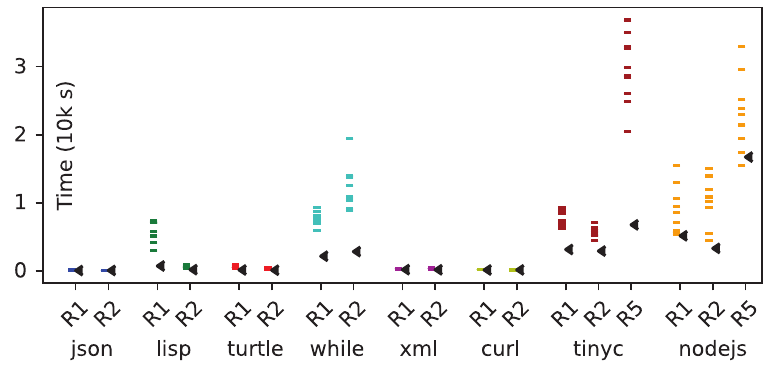}
    \caption{Runtime of Figure~\ref{fig:plot-all-runs} runs; H/R0 runtimes omitted as done on different machines.}
    \label{fig:plot-time}
\end{figure}

For space we only plot their individual results (with the Table~\ref{tab:rerun_Arvada_input_programs}+\ref{tab:perf-results} results for context), i.e., F1~score in Figure~\ref{fig:plot-all-runs} and runtime in Figure~\ref{fig:plot-time}. \toolName{}'s results were stable compared to \arvada{}'s in the sense that across R1, R2, and R5 (except for curl on R1) \toolName{}'s F1 score was better than the average \arvada{} F1 score and sometimes better than every \arvada{} F1 score. Similarly, across all R1, R2, and R5 runs, in the single run where \arvada{} was faster than \toolName{} (js-500 = R5 nodejs) \arvada{}'s F1 score was zero.

Specifically, on the R1 and R2 seeds \toolName{}'s F1 score was better  than all \arvada{} runs for both while and xml.  Across all seed sets, \toolName{} had a better F1~score than at least 9/10 \arvada{} runs for nodejs. Notable outliers are \toolName{}'s while and xml runs on the \arvada{} work's seeds (H). Since the while language uses C-style bracket nesting, \toolName{} does not seem overly overfitted for C-style programming languages.

\subsection{RQ4: Ablation Study}

Here we explore the impact of \toolName{}'s components, using Table~\ref{tab:perf-results}'s experimental setup. We first make Table~\ref{tab:perf-results}'s \arvada{} deterministic (Section~\ref{sec:deterministic-treevada}). For each language the resulting Table~\ref{tab:deterministic-arvada} and following ablation tables show the average of 10 runs. For space we omit standard deviations (all very small or zero). 

Making \arvada{} deterministic yielded several interesting effects. For example, for lisp recall and F1 score are down, runtime is up, parser calls are down, and memory consumption is up---each by about a factor of two. Overall, though, average F1~scores improved while runtimes slowed down. This indicates that some of \arvada{}'s non-deterministic runs got stuck in sub-optimal grammars and thus terminated relatively quickly.

From Table~\ref{tab:deterministic-arvada} to Table~\ref{tab:deterministic+reapply} we add recursive rule application (Section~\ref{sec:solution_reapply}). For most languages this change had little impact on F1 scores. The notable exception are nodejs, tinyc-500, and nodejs-500, which all had lower F1 scores and a faster runtime. Since the initial bracket-implied parse tree are not imposed here and partial merge is used, this version still has bubbles breaking nesting rules. Re-applying such learned nesting-breaking rules does not improve F1 scores (but may get the grammar stuck relatively quickly).

From Table~\ref{tab:deterministic+reapply} to Table~\ref{tab:deterministic+brackets+reapply+all-bubbles} we add bracket-based initial parse trees (Section~\ref{sec:initial-tree}) and ignoring likely string literal contents (Section~\ref{sec:string_literals}). Adding these two features makes the F1 score of lisp jump from 0.34 to 0.99 with a decrease in memory use from 58.7 to only 0.07GB. Additionally, lisp's runtime drops from 11.4k to
0.47k seconds. All 10/10 experiments had improvement in runtime, while languages with more nesting structure benefited the most.

From Table~\ref{tab:deterministic+brackets+reapply+all-bubbles} to Table~\ref{tab:deterministic+brackets+reapply+no partial-merge} we remove partial merges. The F1 scores tend to improve. Especially nodejs's F1 score spiked from 0.09 to 0.56, which indicates that partial merges were responsible for some invalid merges, which essentially blocked learning. 
Finally, from Table~\ref{tab:deterministic+brackets+reapply+no partial-merge} to \toolName{} in Table~\ref{tab:perf-results} we add the new bubble ranking scheme (Section~\ref{sec:bubble-ranking}), which has overall neutral results on F1~scores but overall reduces runtimes.

\begin{table}[h!t]
\caption{Table~\ref{tab:perf-results}'s \arvada{} + deterministic
}
    \begin{center}
    \begin{small}
    
\begin{tabular}{l|rrrr|rrrrr}
    \toprule  
& \multicolumn{1}{c}{r}
& \multicolumn{1}{c}{p}
& \multicolumn{1}{c}{f1}
& \multicolumn{1}{c}{t}
& \multicolumn{1}{c}{t\textsubscript{O}} 
& \multicolumn{1}{c}{t\textsubscript{B}} 
& \multicolumn{1}{c}{t\textsubscript{S}}
& \multicolumn{1}{c}{q[k]}
& \multicolumn{1}{c}{m}\\
    \midrule
json	&.98	&1.0	&.99	&.09	&.06	&.01	&.02	&18	&.05	\\
lisp	&.21	&1.0	&.34	&11.5	&.13	&8.6	&2.6	&14	&58.7	\\
turtle	&1.0	&1.0	&1.0	&.27	&.07	&.02	&.15	&16	&.06	\\
while	&1.0	&1.0	&1.0	&13.7	&.38	&1.5	&8.8	&35	&.38	\\
xml	&1.0	&1.0	&1.0	&.42	&.07	&.04	&.19	&19	&.08	\\
\midrule
curl	&.87	&.75	&.80	&.14	&.11	&.00	&.01	&25	&.05	\\
tinyc	&.81	&.60	&.69	&9.5	&.82	&1.7	&4.0	&141	&.66	\\
nodejs	&.05	&.41	&.08	&17.6	&11.7	&2.5	&1.9	&420	&.73	\\
\midrule
c-500	&.68	&.68	&.68	&52.3	&6.4	&8.3	&22.1	&360	&1.6	\\
js-500	&.40	&.29	&.33	&51.2	&39.1	&8.5	&2.6	&692	&9.5	\\
    \bottomrule      
    \end{tabular}
\label{tab:deterministic-arvada}
    
\end{small}
\end{center}
\end{table}

\begin{table}[h!t]
\caption{Table~\ref{tab:deterministic-arvada} + ReApply learnt rules; 
rc~=~reApply count
}
\begin{small}
    \begin{center}
\begin{tabular}{l|rrrr|rrrrrr}
    \toprule
& \multicolumn{1}{c}{r}
& \multicolumn{1}{c}{p}
& \multicolumn{1}{c}{f1}
& \multicolumn{1}{c}{t}
& \multicolumn{1}{c}{t\textsubscript{O}} 
& \multicolumn{1}{c}{t\textsubscript{B}} 
& \multicolumn{1}{c}{t\textsubscript{S}}
& \multicolumn{1}{c}{q[k]}
& \multicolumn{1}{c}{m}
& rc\\
    \midrule
json	&.98	&1.0	&.99	&.10	&.06	&.01	&.02	&20	&.05	&0	\\
lisp	&.21	&1.0	&.34	&11.4	&.13	&8.5	&2.6	&15	&58.7	&1	\\
turtle	&1.0	&1.0	&1.0	&.32	&.09	&.02	&.17	&17	&.06	&0	\\
while	&1.0	&1.0	&1.0	&14.4	&.40	&1.6	&9.2	&38	&.39	&0	\\
xml	&1.0	&1.0	&1.0	&.47	&.08	&.04	&.21	&21	&.08	&3	\\
\midrule
curl	&.83	&.77	&.80	&.15	&.12	&.00	&.01	&27	&.05	&1	\\
tinyc	&.81	&.60	&.69	&9.77	&.91	&1.6	&4.2	&152	&.67	&4	\\
nodejs	&.01	&.44	&.03	&9.71	&6.3	&2.2	&.76	&210	&1.6	&6	\\
\midrule
c-500	&.66	&.62	&.64	&48.2	&9.4	&6.8	&18.7	&480	&1.7	&1	\\
js-500	&.00	&.73	&.00	&45.2	&31.9	&9.6	&2.6	&548	&10.0	&13	\\
    \bottomrule      
    \end{tabular}
\label{tab:deterministic+reapply}
\end{center}
\end{small}
\end{table}

\begin{table}[h!t]
\caption{Table~\ref{tab:deterministic+reapply}  
+ Initial bracket tree 
+ Ignore string literals
}
    \begin{center}
    \begin{small}
\begin{tabular}{l|rrrr|rrrrrr}
    \toprule
& \multicolumn{1}{c}{r}
& \multicolumn{1}{c}{p}
& \multicolumn{1}{c}{f1}
& \multicolumn{1}{c}{t}
& \multicolumn{1}{c}{t\textsubscript{O}} 
& \multicolumn{1}{c}{t\textsubscript{B}} 
& \multicolumn{1}{c}{t\textsubscript{S}}
& \multicolumn{1}{c}{q[k]}
& \multicolumn{1}{c}{m}
& \multicolumn{1}{c}{rc}\\
    \midrule
json	&.98	&.90	&.94	&.05	&.04	&.00	&.00	&10	&.03	&2	\\
lisp	&.98	&1.0	&.99	&.47	&.32	&.00	&.07	&44	&.07	&2	\\
turtle	&1.0	&1.0	&1.0	&.16	&.09	&.01	&.04	&18	&.04	&0	\\
while	&.98	&1.0	&.99	&3.7	&.65	&.29	&1.8	&75	&.21	&1	\\
xml	&1.0	&1.0	&1.0	&.19	&.08	&.03	&.05	&16	&.07	&2	\\
\midrule
curl	&.58	&.84	&.69	&.14	&.12	&.00	&.01	&24	&.04	&1	\\
tinyc	&.88	&.43	&.58	&3.6	&.85	&.19	&1.4	&200	&.20	&6	\\
nodejs	&.05	&.54	&.09	&7.1	&6.3	&.03	&.38	&183	&.08	&8	\\
\midrule
c-500	&.66	&.74	&.69	&16.3	&2.6	&1.2	&7.5	&459	&.60	&4	\\
js-500	&.56	&.35	&.43	&12.8	&10.9	&.11	&.77	&298	&.33	&13	\\
    \bottomrule      
    \end{tabular}
\label{tab:deterministic+brackets+reapply+all-bubbles}
\end{small}
\end{center}
\end{table}

\begin{table}[h!t]
\caption{Table~\ref{tab:deterministic+brackets+reapply+all-bubbles}  
+ Remove partial merge
}
    \begin{center}
    \begin{small}
\begin{tabular}{l|rrrr|rrrrrr}
    \toprule
& \multicolumn{1}{c}{r}
& \multicolumn{1}{c}{p}
& \multicolumn{1}{c}{f1}
& \multicolumn{1}{c}{t}
& \multicolumn{1}{c}{t\textsubscript{O}} 
& \multicolumn{1}{c}{t\textsubscript{B}} 
& \multicolumn{1}{c}{t\textsubscript{S}}
& \multicolumn{1}{c}{q[k]}
& \multicolumn{1}{c}{m}
& \multicolumn{1}{c}{rc}\\
    \midrule
json	&.90	&.98	&.93	&.04	&.03	&.00	&.00	&10	&.03	&0	\\
lisp	&1.0	&1.0	&1.0	&.60	&.36	&.00	&.11	&47	&.07	&0	\\
turtle	&1.0	&.67	&.80	&.27	&.10	&.01	&.09	&22	&.04	&0	\\
while	&1.0	&1.0	&1.0	&2.3	&.20	&.17	&1.2	&33	&.13	&0	\\
xml	&1.0	&1.0	&1.0	&.16	&.06	&.03	&.04	&13	&.07	&0	\\
\midrule
curl	&.89	&.90	&.89	&.13	&.11	&.01	&.01	&26	&.05	&1	\\
tinyc	&.87	&.56	&.68	&2.8	&.46	&.18	&1.2	&136	&.18	&0	\\
nodejs	&.44	&.76	&.56	&5.7	&5.1	&.02	&.30	&131	&.07	&2	\\
\midrule
c-500	&.46	&.97	&.62	&12.3	&1.1	&.60	&6.4	&235	&.35	&0	\\
js-500	&.48	&.57	&.52	&17.6	&14.7	&.11	&1.2	&410	&.34	&0	\\
    \bottomrule      
    \end{tabular}
\label{tab:deterministic+brackets+reapply+no partial-merge}
\end{small}
\end{center}
\end{table}

\section{Threats to Validity}

We briefly summarize key threats to internal and external validity.

\textbf{Threats to external validity:} 
Often grammar inference tools show mixed performance with different seeds~\cite{Glade22Replication}. Also relying on hand-crafted small toy programs as seed does not replicate real-world situations. To overcome these challenges, we randomly sample seed sets R1, R2, R5 for our experiments.

\textbf{Threats to internal validity:}
\toolName{}'s pre-structuring may fail if brackets serve other language-specific purposes than nesting. 
For example, in \texttt{xml} names enclosed in angle brackets \term{<} \term{>} are used for nesting, whereas in Java or C++ these are mainly used for relational or bit manipulation operators. We have carefully chosen only three bracket types for pre-structuring parse tree as these three are the most commonly used for nesting~\cite{van2019lightweight}. For terminal expansion, synthesizing regular expressions for the terminals would give a more robust grammar, which is future work.

\section{Related Work}

Grammatical inference is important as the inferred grammar can serve many tasks~\cite{stevenson2014survey} when the language's golden grammar is unknown. 
Rather than probabilistic learning~\cite{valiant1984theory, li1991learning}, \textit{active learning}~\cite{angluin1981note} is a good fit as a parser can serve as a \textit{minimally adequate teacher} (\textsc{Mat} or oracle). Even as a black-box the \textsc{Mat}/oracle can answer membership queries. 
Recent work~\cite{Glade17PLDI, REINAM2019FSE, arvada21ASE} follows this setting to infer grammars. Following is other related work.

\subsection{Linguistics}

The linguistics community has developed several negative~\cite{angluin1991won,eyraud2007lars,de1997characteristic} and positive results on grammar inference in a wide variety of settings, i.e., for various grammar classes (including context-free), oracles, and availability of additional kinds of input~\cite{Sakakibara97Recent}. GRIDS~\cite{langley2000learning} starts with flat parse trees and iteratively bubbles and merges rules. Maybe most closely related to \arvada{} is applying a GRIDS-like approach iteratively~\cite{Zuidema02How} and thus sampling new inputs from an updated grammar.
Maybe most closely related to \toolName{} are Sakakibara's techniques for inferring a subclass of context-free grammars when given positive examples with their complete (but unlabeled) parse trees plus a parser-like oracle and a grammar equivalence oracle~\cite{sakakibara1990learning,sakakibara1992efficient}.

\balance

We observe that a program's bracket-implied nesting structure can be captured by a \texttt{Dyck} language~\cite{berstel2002balanced}, i.e., a context free language comprising of only balanced brackets. \toolName{}'s pre-structured parse trees can thus be seen as instances of the Dyck language $D_3$ on the alphabet \term{(}, \term{)}, \term{[}, \term{]}, \term{\{}, \term{\}}. An interesting property is that a Dyck language captures the ``non-regular essence'' of a context-free language. Specifically, the Chomsky--Sch\"{u}tzenberger representation theorem~\cite{chomsky} says any context-free language can be mapped to the intersection of a Dyck language and a regular language. 

For \toolName{}, this pre-structuring is just a startup heuristic. For example, in subsequent steps \toolName{} may discover additional balanced structures defined by some other opening and closing terminal pair (which taken together may then be represented by a $D_4$ language). Similarly, \toolName{} may not be able to merge rules in the initial parse trees.

\subsection{Deep Learning}

Several reports have been negative on using deep learning for inference of a context-free grammar (CFG)~\cite{yellin2021synthesizing,sennhauser2018evaluating}. RNNs lack the ability to learn concrete hierarchical rules, leading to a decline in generalization with increasing input length and recursion depth~\cite{bernardy2018can,yu2019learning}. LSTMs learn statistical approximations, not a deterministic rule-based solution~\cite{sennhauser2018evaluating}. Even the state of the art attention based Seq2Seq model struggles to understand CFGs~\cite{yu2019learning}. 
\arvada{} showed significantly better precision than an LSTM-based model (measuring recall is not possible). Deep learning's under-performance may be due to solely relying on (many) input samples. \glade{}, \arvada{}, and \toolName{} use \textit{active learning}~\cite{angluin1981note}, where an additional oracle guides the learning process. 
Another significant deep learning limitation is that it does not produce an explicit grammar, which makes it also hard to measure recall and F1 score.

\subsection{Grey-box Grammar Inference}

The grey-box grammar inference approaches don't make use of the entire parser source code. \textit{GRIMOIRE}~\cite{blazytko2019grimoire} is a grey-box fuzzing tool that makes use of the parser's coverage information. \textit{GRIMOIRE} synthesizes a grammar like structure of the inputs while fuzzing.

\subsection{White-box Grammar Inference}

White-box grammar inference approaches utilize the parser source code to extract input grammars that follow the structure of the input. Lin et al.~\cite{lin2008deriving,lin2009reverse} proposed the first white-box method that recovers parse trees from inputs using static and dynamic analysis. \textit{Autogram}~\cite{Autogram2016ASE}, introduced by Höschele et al, adopts another white-box approach that tracks the dynamic data flow between variables of the program to infer an approximate context-free grammar. \textit{Mimid}~\cite{gopinath20mining} by Gopinath et al. infers a grammar by leveraging dynamic control flow and tracking input character access across parser locations.

\section{Conclusions}

Black-box context-free grammar inference is a hard problem as in many practical settings it only has access to a limited number of example programs. The state-of-the-art approach \arvada{} heuristically generalizes grammar rules starting from flat parse trees and is non-deterministic to explore different generalization sequences. We observe that many of \arvada{}'s generalization steps violate common language concept nesting rules. We thus propose to pre-structure input programs along these nesting rules, apply learnt rules recursively, and make black-box context-free grammar inference deterministic. The resulting \toolName{} yielded faster runtime and higher-quality grammars in an empirical comparison. The \toolName{} source code, scripts, evaluation parameters, and training data are open-source and publicly available.

\begin{acks}
The authors acknowledge the Texas Advanced Computing Center (TACC) at The University of Texas at Austin for providing HPC resources that have contributed to the research results reported within this paper (\url{http://www.tacc.utexas.edu}). 
Christoph Csallner has a potential research conflict of interest due to a financial interest with Microsoft and The Trade Desk. A management plan has been created to preserve objectivity in research in accordance with UTA policy. This material is based upon work supported by the National Science Foundation (NSF) under Grant No. 1911017 and a gift from MathWorks.
\end{acks}

\bibliographystyle{ACM-Reference-Format}
\bibliography{main}

\end{document}